\documentclass{aa}
\usepackage{graphicx}
\usepackage{txfonts}
\usepackage{hyperref}
\hypersetup{
    colorlinks=true,
    linkcolor=blue,
    filecolor=magenta,
    urlcolor=cyan,
    citecolor=blue,
    pdftitle={IEF Topology},
    pdfpagemode=FullScreen,
    }

\usepackage{natbib,color}
\usepackage{amsmath,amssymb}
\usepackage[nice]{nicefrac} 
\usepackage{textcomp} 
\usepackage{gensymb} 
\usepackage[normalem]{ulem} 

\begin{document}

   \title{The magnetic topology of the inverse Evershed flow}


   \author{A. Prasad\inst{1,2,3}
          \and
          M. Ranganathan\inst{4,5}
         \and
         C. Beck \inst{6}
         \and
         D. P. Choudhary \inst{4}
         \and
         Q. Hu \inst{3,7}
          }

   \institute{Rosseland Centre for Solar Physics, University of Oslo, Postboks 1029 Blindern, 0315 Oslo, Norway \\
   \email{avijeet.prasad@astro.uio.no}
    \and
   Institute of Theoretical Astrophysics, University of Oslo, Postboks 1029 Blindern, 0315 Oslo, Norway
    \and
   Center for Space Plasma \& Aeronomic Research,
                The University of Alabama in Huntsville,
                Huntsville, Alabama 35899, USA
    \and
    Department of Physics \& Astronomy, California State University, Northridge, CA 91330-8268, USA
    \and
    Institute for Particle and Astrophysics, ETH, Zürich, Switzerland 8049
    \and
    National Solar Observatory (NSO), 3665 Discovery Drive, Boulder, CO 80303, USA
    \and
    Department of Space Science, The University of Alabama in Huntsville, Huntsville, AL 35899, USA
             }


  \abstract
   {The inverse Evershed flow (IEF) is a mass motion towards sunspots at chromospheric heights.}
   {We combined high-resolution observations of NOAA 12418 from the Dunn Solar Telescope and vector magnetic field measurements from the Helioseismic and Magnetic Imager (HMI) to determine the driver of the IEF.}
   {We derived chromospheric line-of-sight (LOS) velocities from spectra of H$\alpha$ and \ion{Ca}{ii} IR. The HMI data were used in a non-force-free magnetic field extrapolation to track closed field lines near the sunspot in the active region.
We determined their length and height, located their inner and outer foot points, and derived flow velocities along them.}
   {The magnetic field lines related to the IEF reach on average a height of 3\,Mm over a length of 13\,Mm. The inner (outer) foot points are located  at 1.2 (1.9) sunspot radii. The average field strength difference $\Delta B$ between inner and outer  foot points is +400\,G. The temperature difference $\Delta T$ is anti-correlated with $\Delta B$ with an average value of -100\,K.
The pressure difference $\Delta p$ is dominated by $\Delta B$ and is primarily positive with a driving force towards the inner foot points of 1.7\,kPa on average. The velocities predicted from $\Delta p$ reproduce the LOS velocities of 2--10\,km s$^{-1}$ with a square-root dependence.}
   {We find that the IEF is driven along magnetic field lines connecting network elements with the outer penumbra by a gas pressure difference that results from a difference in field strength as predicted by the classical siphon flow scenario.}

   \maketitle
%

\section{Introduction}
\label{s:intro}
The chromospheric inverse Evershed flow \citep[IEF;][]{evershed1909a} transports material into sunspots along magnetic field lines that connect the boundary of the moat cell with the outer penumbra. These closed loops and the flows along them have a lifetime of a few tens of miutes to more than 1\,h \citep{georgakilas+christopoulou2003,beck+choudhary2020}. The flow velocities of about $2-9$\,km\,s$^{-1}$ \citep{dialetis+etal1985,dere+etal1990,beck+etal2020} carry the material along loops that ascend to a height of a few Mm \citep{maltby1975,beck+etal2020}. Flows in dark super-penumbral fibrils were found to be faster than those in bright ones, with fluctuations on timescales of about 25\,mins \citep{georgakilas+etal2003,georgakilas+christopoulou2003}. At the inner foot points in the penumbra, the magnetic field lines return to the photosphere with an average angle of about 65$^\circ$ to the local vertical \citep{haugen1969,beck+choudhary2019}. As the flow speed at the inner foot point exceeds the photospheric sound speed, the flow terminates in a stationary shock front that heats the lower chromosphere \citep{thomas+montesinos1991,beck+etal2014,choudhary+beck2018}.

The magnetic and thermodynamic properties and the lifetime of IEF channels comply with a siphon flow scenario \citep{meyer+schmidt1968,thomas1988,montesinos+thomas1997}, where a stationary flow can be driven along magnetic fields lines (MFLs) that have two foot points (FPs) of different magnetic field strength. If the two FPs have the same total pressure, the difference in field strength causes a difference in gas pressure as the necessary mechanical driving force of the flow.  Observational evidence of siphon flows in the solar atmosphere was reported by, e.g., \citet{ruedi+etal1992}, \citet{uitenbroek+etal2006} and \citet{bethge+etal2012}, but flows along MFLs that connect FPs of unequal field strength can also be driven by other physical processes \citep{sigwarth+etal1998}. A major problem to diagnose siphon flows in solar observations is to unequivocally identify the two FPs to determine their field strength, which requires knowledge of the magnetic connectivity.

Narrow filaments in the photosphere and fibrils in the chromosphere are known to trace MFLs connecting the central regions of sunspots with their surroundings \citep{frazier1972,zirin1972,schad+etal2013,beck+choudhary2019}, but especially for IEF channels it is usually impossible to locate their outer FPs, as their signature in velocity and intensity fades into the background \citep[see, e.g.,][]{beck+etal2020}.
A more direct tool to derive the magnetic connectivity of sunspots or whole active regions  are magnetic field extrapolations that utilize photospheric (vector) magnetograms to infer the coronal magnetic field under certain assumptions. For example, the assumption that the magnetic pressure dominates over the plasma pressure (plasma $\beta << 1$) allows one to neglect all non-magnetic forces and assume the Lorentz-force to be zero. This approach leads to force-free fields, which have been widely used in the solar community \citep[e.g.,][]{wiegelmann2008jgra,wiegelmann&sakurai2012lrsp}. However, it was pointed out by \citet{gary2001soph} that the plasma $\beta$ can be of the order of unity in solar photosphere, where the magnetic field measurements are taken, thus emphasizing the need for a different approach that incorporates non-force-free effects. One such alternative is the non-force-free-field (NFFF) extrapolation technique  \citep{hu08a,hu08b,hu10} based on the principle of the minimum energy dissipation rate \citep{bhattacharyya07}, where the magnetic field is expressed as the superposition of one potential field and two (constant $\alpha$) linear force-free fields with distinct $\alpha$ parameters. Such NFFF extrapolations have been used in many recent studies to model flaring active regions and coronal jets \citep{nayak+2019apj,liu+2020ApJ,yalim+2020ApJ,prasad+2020ApJ}. Interested readers are referred to appendix A of \citet{liu+2020ApJ}, which provides a detailed discussion on the applicability of the NFFF method.
\begin{figure*}[h!]
\resizebox{\hsize}{!}{\includegraphics{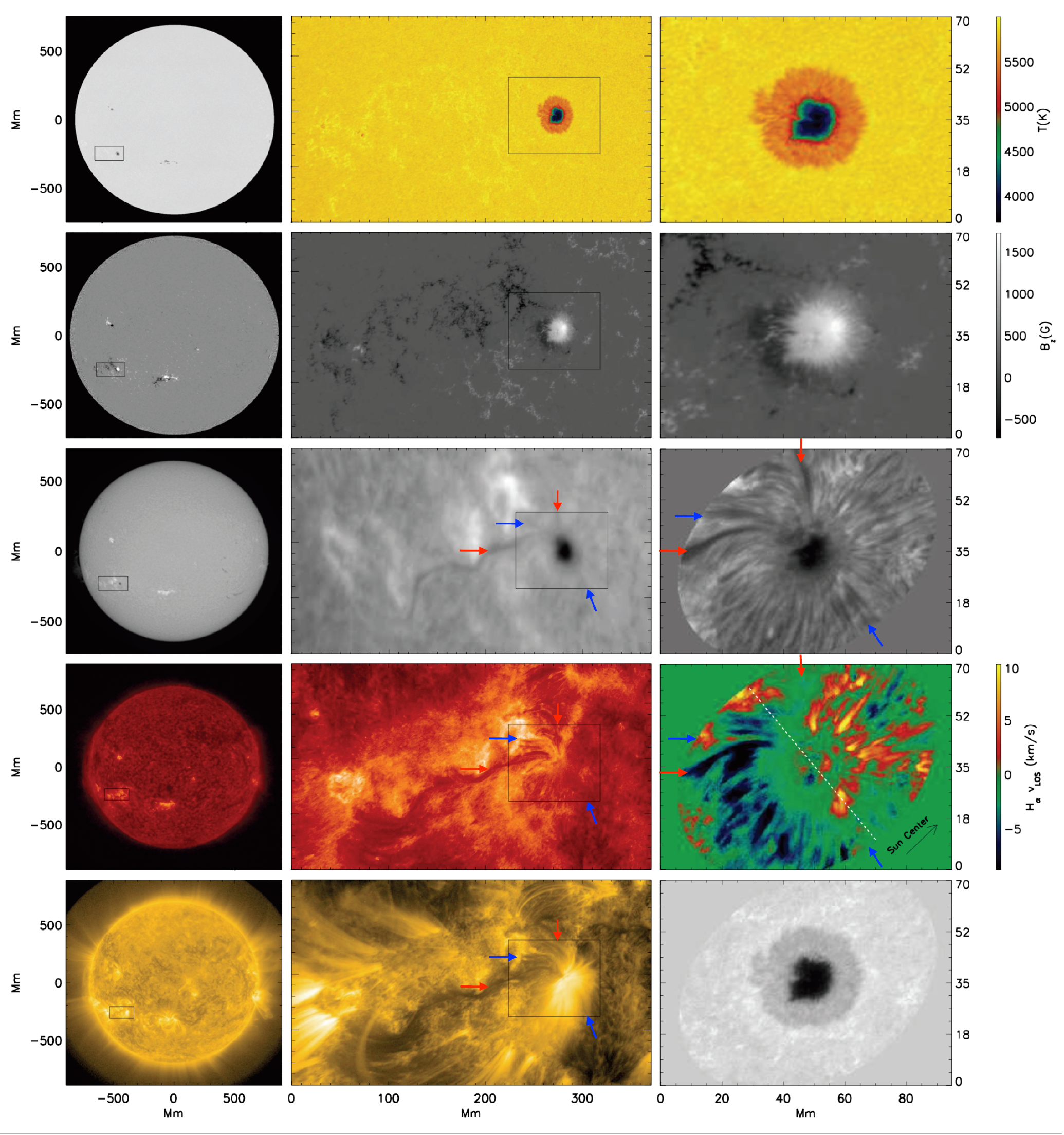}}
     \caption{Overview figure showing the location of AR NOAA 12418 on 16 September 2015. Left column, top to bottom: full-disk images in HMI $I_c$, $B_z$, GONG H$\alpha$, AIA 304\,{\AA}, and AIA 171\,{\AA}. Middle column: magnification of the black rectangle (371\,Mm $ \times$ 186\,Mm) in the left column. The HMI $I_c$ has been exchanged for the derived temperature instead. Right column: magnification of the black rectangle marked in the middle column (94\,Mm $ \times$ 70\,Mm). The lower three panels show the H$\alpha$ line-core intensity and velocity, and the continuum intensity from the IBIS high-resolution spectra. The white dashed line in the H$\alpha$ velocity map marks the symmetry line of the sunspot. The GONG H$\alpha$ image in the middle column is not de-projected in contrast to all other panels in the right two columns.  The red and blue arrows mark examples of H$\alpha$ filaments as opposite to IEF channels.} \label{fig_overview}
   \end{figure*}

In this paper, we investigate the magnetic field properties of  IEF channels using a combination of high-resolution observations of chromospheric spectral lines to trace the flow channels and an NFFF extrapolation to determine the magnetic connectivity.
The use of extrapolated field lines allows us to identify the foot points and determine the heights of flow channels that connect the opposite-polarity patches in the magnetograms. As the field gradient and the height of the flow channels play a crucial role in driving the siphon flows, we employ these quantities to study properties such as the strength of mass motions in inverse Evershed flows.  We use the same definition as in \citet{beck+etal2020} that super-penumbral, roughly radially oriented elongated fibrils that connect the penumbra with the super-penumbral boundary and that exhibit a significant flow velocity constitute IEF channels.

Section \ref{secobs} describes the observations used. Our analysis methods are explained in Section \ref{secana} and the Appendices A to D. The analysis results are given in Section \ref{secres}. Section \ref{secdisc} discusses the findings, while Section \ref{secconcl} provides our conclusions.

\section{Observations}\label{secobs}

We observed the decaying active region (AR) NOAA 12418 on 16 September 2015 with the Interferometric BIdimensional Spectrometer \citep[IBIS;][]{cavallini2006,reardon+cavallini2008} and the Facility InfraRed Spectropolarimeter \citep[FIRS;][]{jaeggli+etal2010} at the Dunn Solar Telescope \citep[DST;][]{dunn1969, dunn+smartt1991}. The field of view (FOV) covered the isolated leading sunspot of the AR, located at about $x,y = -500^{\prime\prime},-340^{\prime\prime}$ at a heliocentric angle of 43$^\circ$. The DST observations are described in detail in \citet{beck+choudhary2020}, so we will only give a short summary here.

IBIS was used to sequentially obtain 400 spectral scans of the two chromospheric spectral lines of H$\alpha$ at 656\,nm and \ion{Ca}{ii} IR at 854.2\,nm with about 30 wavelength points each between 14:42 and 15:56 UT. The circular IBIS FOV had a diameter of 95$^{\prime\prime}$ at a spatial sampling of 0\farcs095\,pixel$^{-1}$. For the current study, we only used the six spectral scans between 14:48 and 15:48 UT that were simultaneous to full-vector observations with the Helioseismic and Magnetic Imager \citep[HMI;][]{scherrer+etal2012} at the 12-min cadence of the latter. The IBIS and HMI data were complemented with full-disk images at 171, 304, and 1700\,{\AA} from the Atmospheric Imaging Assembly \citep[AIA;][]{lemen+etal2012} on-board the Solar Dynamics Observatory \citep[SDO;][]{pesnell+etal2012} and H$\alpha$ images from the Global Oscillation Network Group \citep[GONG;][]{harvey+etal1996}.

Figure \ref{fig_overview} shows an overview of the AR in different quantities. The panels in the middle column correspond to the box used for the magnetic field extrapolation, while the rightmost column corresponds to the IBIS FOV after de-projection. The AR was located in the south-east quadrant of the Sun. The leading positive polarity contained a single, round sunspot, while the trailing negative polarity had decayed to more diffuse plage regions. There was no significant amount of magnetic flux of negative polarity to the west of the sunspot.

The H$\alpha$, and AIA 171 and 304\,{\AA} images show a large filament of about 200\,Mm length towards the east and a smaller filament of 75\,Mm length towards the north starting from the outer penumbra of the sunspot. These two filaments are located above polarity inversion lines and are marked with red arrows in Figure \ref{fig_overview}. We marked two more shorter filaments with blue arrows in Figures \ref{fig_overview} and \ref{f:selected}, where the relation to a neutral line is less clear. All of those filaments differ from IEF channels by having a much stronger absorption, a larger lateral width, an extended velocity signature and a longer lifetime without any obvious changes over the 1-hr duration of the observations (see also Figure 12 of \citeauthor{beck+choudhary2020} \citeyear{beck+choudhary2020}). They are likely to correspond to the topmost layer of the solar chromosphere that attains enough mass and hence opacity to cause such absorption \citep{leenaarts+etal2015}.

The AIA 171\,{\AA} image shows some closed loops from the sunspot to the trailing plage, but no obvious loops towards the west. The symmetry line of the sunspot with roughly zero line-of-sight (LOS) velocities due to the projection effects on the LOS makes an angle of about 45$^\circ$ going from the north-east to the south-west. The IEF channels are clearly seen in the H$\alpha$ line-core intensity and velocity maps, with a roughly even distribution in azimuth around the sunspot. The downflow patches on the center side are more pronounced and isolated than those on the limb side, where the end points of the IEF channels to some extent fall on the neutral line of LOS velocities.

\section{Data analysis}\label{secana}
The majority of the data analysis and the alignment of the high-resolution and full-disk data follow common procedures and are  described in detail in the Appendices \ref{appendix_a} to \ref{appendix_d}. We only summarize the steps and their outcome here.

From the HMI data, we retrieved the photospheric LOS velocities while compensating the solar rotation across the large FOV (Appendix ~\ref{app_photvelo}). Chromospheric LOS velocities were derived from the H$\alpha$ and \ion{Ca}{ii} IR 854\,nm spectra from IBIS using a bi-sector method. The average velocities across the IBIS FOV were set to zero. Both lines were found to yield very similar values and can thus be used synonymously (Appendix~\ref{app_chromvelo}). The HMI continuum intensity was normalized to unity in the quiet Sun and then converted to temperature using the Planck function (Appendix~\ref{appendix_b}).
The HMI vector magnetograms were extrapolated using the NFFF method to retrieve the chromospheric and coronal magnetic field in a three-dimensional (3D) volume (Appendix~\ref{appendix_c}). The ground-based high-resolution and space-based full-disk data were de-projected to correct for the geometrical foreshortening and subsequently aligned to each other with the HMI continuum intensity image as the reference (Appendix~\ref{appendix_d}). As an estimate of the true velocity magnitude, the chromospheric LOS velocities were de-projected onto the magnetic field vector in the magnetic field extrapolation at a height of 1\,Mm (Appendix~\ref{app_deprovelo}) to remove the LOS projection effects assuming field-aligned flows. The data analysis steps that are less common are described in the following two sections.
\begin{table}
\caption{HSRA values at a few selected heights.}
\begin{tabular}{ccccc}
z & $\log\tau$ & T & $\rho_{\rm gas}$ & $p$\cr
km & -- & K & $10^{-7}$\,g\,cm$^{-3}$&kPa\cr\hline\hline
-77 & 1.3 &9390& 3.24  &20.01\cr
-25 & 0.3 & 7140& 3.36 &15.40\cr
11& -0.1 & 6200 & 3.06&12.17\cr
23& -0.2 &6035&2.89   &11.20\cr
35&-0.3  &5890&2.70   &10.21\cr
49&-0.4  &5765&2.50   &9.24\cr
63&-0.5  &5650&2.29   &8.31 \cr
138&-1   &5160& 1.38  &4.56\cr
283 & -2 & 4660& 0.43 &1.27\cr
\end{tabular}
\label{table:HSRA}
\end{table}
\begin{figure*}
\resizebox{17.6cm}{!}{\includegraphics{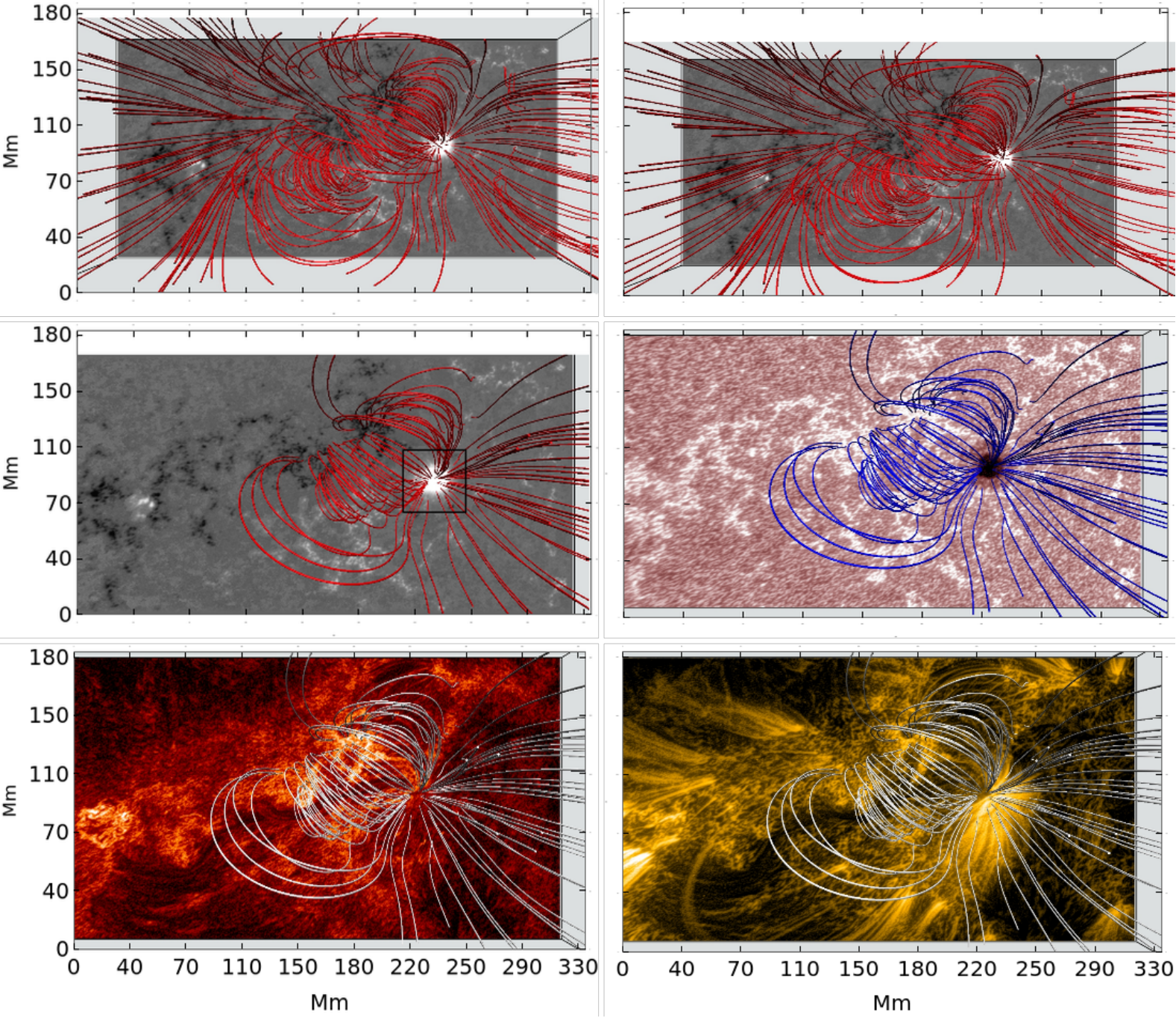}}
\caption{Large-scale magnetic field topology of AR 12418 on 2015 Sep 16 from NFFF extrapolations. Top row: full domain at 14:48 and 15:48 UT on top of $B_z$ with open and closed MFLs. Middle (bottom) row: MFLs originating in the sunspot at 14:48 UT on top of $B_z$ and AIA 1700\,{\AA} (AIA 304\,{\AA}  and AIA 171\,{\AA}). The black square in the middle left panel indicates the area of the initial seed points for the identification of closed MFLs. The FOV of the top row is slightly zoomed out to show the large-scale connectivity.} \label{f:nfff}
\end{figure*}
\subsection{Pressure balance equation}
In the simplest approximation of magneto-hydrostatic equilibrium for the ionized plasma of the solar atmosphere in the presence of magnetic fields, the total pressure $p_{\rm tot}$ is given by the addition of the magnetic pressure $p_{\rm mag}$ and gas pressure $p_{\rm gas}$ by \citep[e.g.,][]{steiner+etal1986,thomas1988}

\begin{equation}
p_{\rm tot} = p_{\rm mag} + p_{\rm gas} = \frac{B^2}{2\,\mu_0} + \frac{\rho}{\mu}\, R\, T\,, \label{eq_totpres}
\end{equation}

with the magnetic field $B$ in Tesla T, the magnetic permeability constant $\mu_0 = 1.26\times 10^{-6}$\,H\,m$^{-1}$, the gas density $\rho$ in kg\,m$^{-3}$, the mean molecular weight of the Sun $\mu = 1.3\,$g\,mol$^{-1}$, the universal gas constant $R = 8.314\,$kg\,m$^2\,$s$^{-2}$\,K$^{-1}$\,mol$^{-1}$ and the gas temperature $T$ in K.

Two points 1 and 2 in the solar atmosphere that are connected by MFLs or that are in direct proximity have magnetic and gas pressure differences $\Delta p_{\rm mag}$ and $\Delta p_{\rm gas}$ of

\begin{equation}
\Delta p_{\rm mag} = \frac{B_1^2 - B_2^2}{2\,\mu_0}\,\mbox{\,\,and\,\,}\,\Delta p_{\rm gas} = \frac{R\,(\rho_1\,T_1 - \rho_2\,T_2)}{\mu}\,, \label{eq_inoutp}
\end{equation}

with the magnetic field strengths $B_i$, temperatures $T_i$ and densities $\rho_i$ with $i=1,2$.

For spatially separated points, a difference in gas pressure can drive a mass flow along a connecting field line. For $B_1 > B_2$ and total pressure equilibrium, $p_{\rm gas}^1$ is smaller than $p_{\rm gas}^2$, and the flow moves from the location with lower to higher field strength, which is called a siphon flow \citep{meyer+schmidt1968,cargill+priest1980}. For $T_1 < T_2$ at equal density, the flow goes from the point with higher to the one with lower temperature regardless of total pressure equilibrium.

As it is not instantly clear if total pressure equilibrium is valid over large distances, we defined additionally the effective gas pressure difference $\Delta p_{\rm eff}$ by

\begin{equation}
\Delta p_{\rm eff} = \frac{B_1^2 - B_2^2}{2\,\mu_0} + \frac{\rho}{\mu}\, R\, (T_2 - T_1)\label{p_effect}
\end{equation}

as the effective driving force of flows along connecting magnetic field lines, assuming $\rho_1 = \rho_2 = \rho$.

If the lateral total pressure equilibrium holds over an extended spatial area, Equation (\ref{eq_totpres}) predicts a linear relation between the temperature $T$ and $B^2$ as

\begin{equation}
T(B^2) = c - \frac{\mu }{2\,\mu_0 R \rho} B^2\,, \label{eq_density}
\end{equation}

where the slope is proportional to $\rho^{-1}$ is the only unknown if $T$ and $B$ are given, while $c$ is proportional to the total pressure.

For a stationary gas flow driven by a pressure difference $\Delta p$, a first-order estimate of the flow speed $v$ can be derived through the kinetic energy density $E_{\rm kin} = \frac{1}{2}\,\rho\,v^2$ as \citep{bethge+etal2012}

\begin{eqnarray}
v(\Delta p) = \sqrt{\frac{2\,\Delta p}{\rho}}\,, \label{speed_eq}
\end{eqnarray}
while the potential energy $E_{\rm pot} = \rho \,g\,h$ allows one to estimate the maximal height that can be attained through

\begin{eqnarray}
h(\Delta p) = \frac{\Delta p}{\rho\,g}\label{height_eq1}
\end{eqnarray}

with the solar surface gravity $g = 273$\,m\,s$^{-2}$.

Finally, equating kinetic and potential energy density gives the maximal height in dependence of the flow speed as

\begin{equation}
h(v) = \frac{v^2}{2 g} \,.\label{height_eq2}
\end{equation}

 Table \ref{table:HSRA} lists the values of formation height, optical depth, temperature, gas density and total pressure in the Harvard Smithsonian Reference Atmosphere \citep[HSRA;][]{gingerich+etal1971} over a height range from -100--300\,km as an example of typical values of thermodynamic properties.

\begin{figure}
\resizebox{8.8cm}{!}{\includegraphics{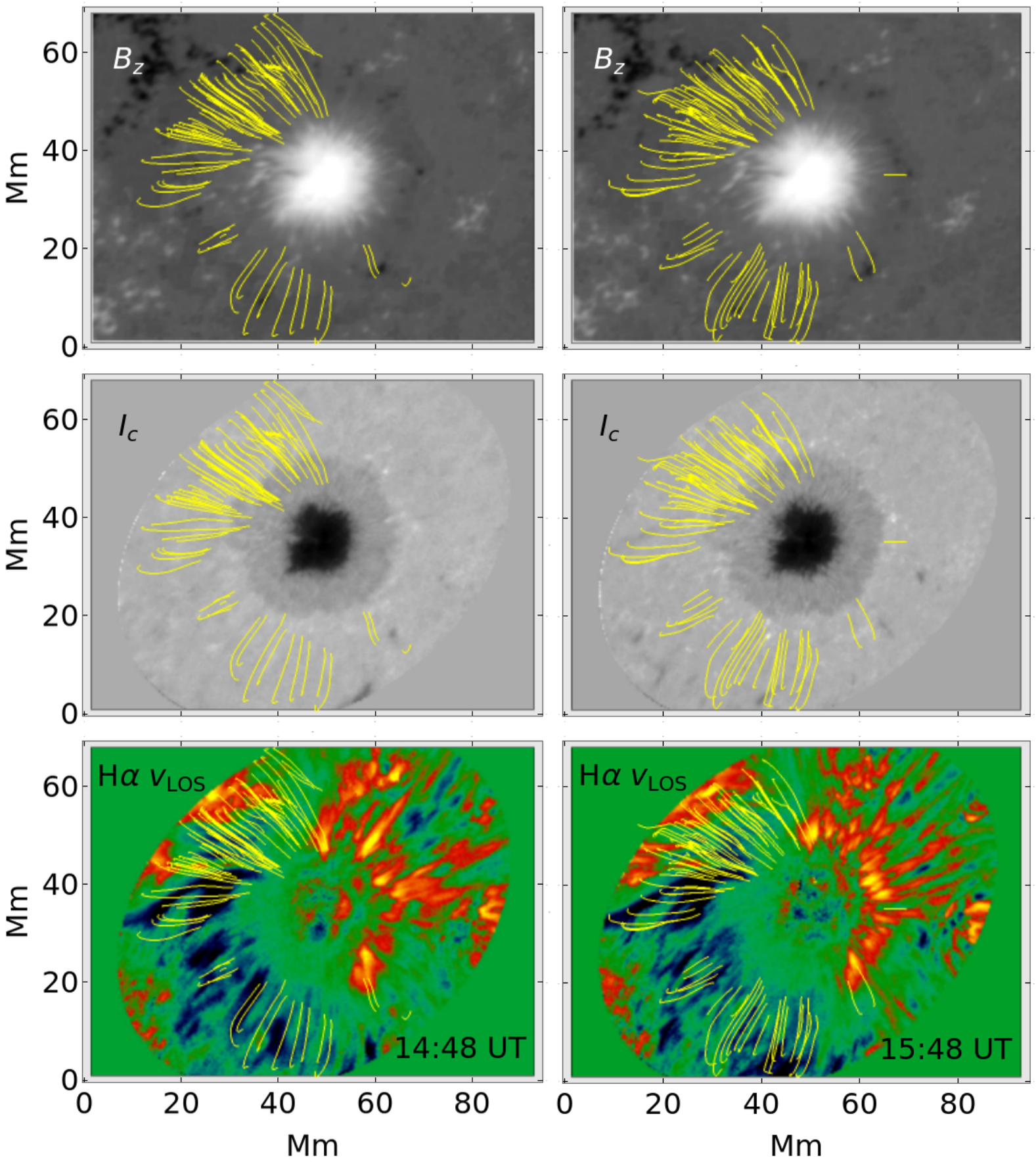}}
\caption{Examples of automatically selected closed MFLs around the sunspot. Left (right) column: at 14:48 (15:48) UT.  Only 1/33 of the MFLs are drawn. The background images show from top to bottom $B_z$, the H$\alpha$ continuum intensity, and the H$\alpha$ LOS velocity.} \label{f:full}
\end{figure}
\subsection{Selection of field lines and IEF channels}
The top row of Figure \ref{f:nfff} shows an overview of the magnetic field extrapolation results for the whole AR for the first and last magnetograms at 14:48 and 15:58 UT, respectively. The global magnetic topology will be described below, but only some MFLs are relevant for the IEF channels. We thus used two subsets of closed MFLs defined in an automatic and a manual way for closer investigation.

\begin{table}
\caption{Fraction of MFLs meeting combined selection criteria.}\label{tab_select}
\begin{tabular}{ccccc}
total & closed & $L > 5.4$\,Mm & $z<7.5\,$Mm & Inside IBIS FOV\cr\hline\hline
22500 &  7103 & 4502 & 3915 & 3154\cr
100\,\% & 32\,\% & 20\,\% & 17\,\% & 14\,\%\cr
\end{tabular}
\end{table}

\begin{figure*}
\resizebox{17cm}{!}{\includegraphics{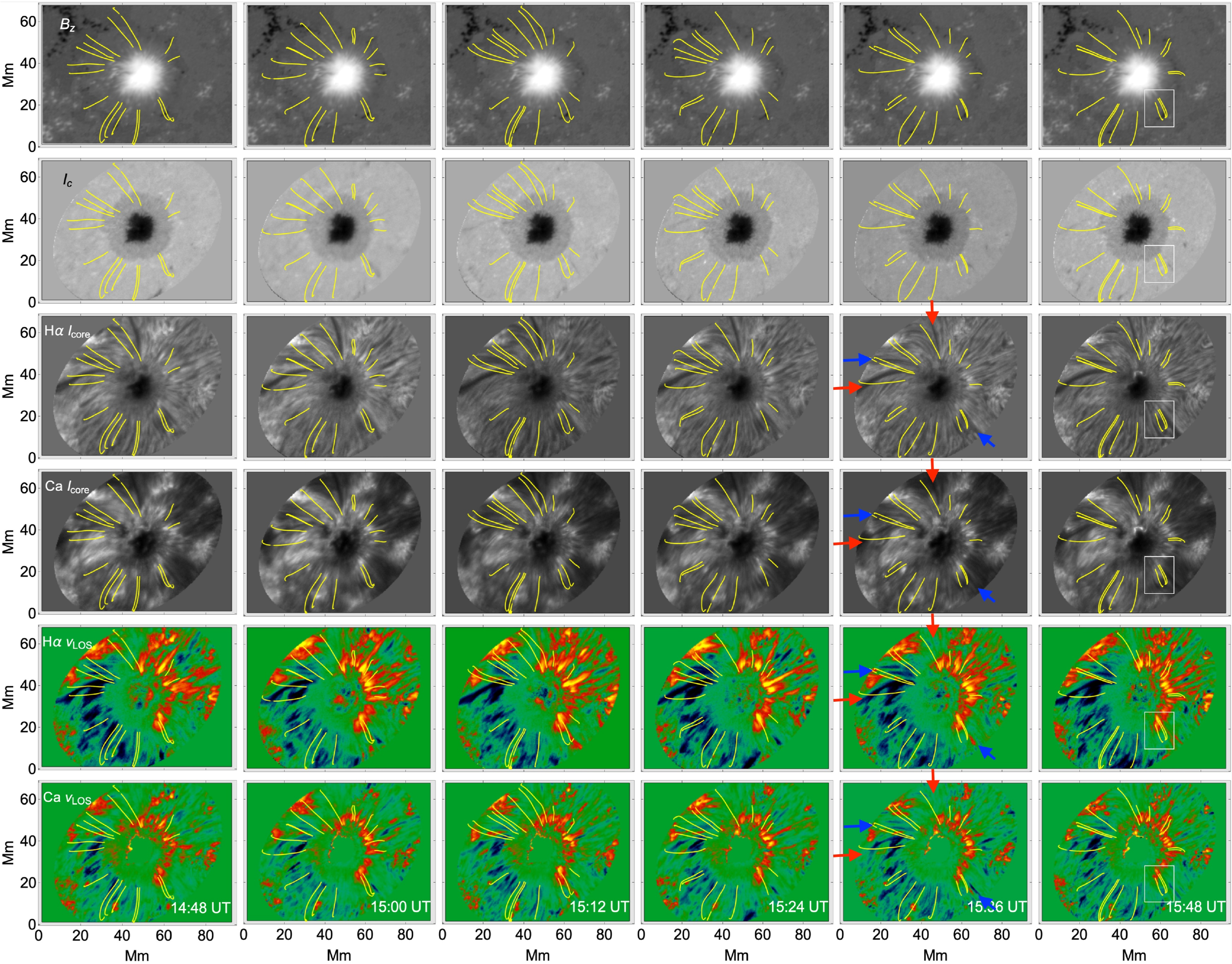}}
\caption{Magnetic field lines of all manually selected IEF channels. Left to right: at 14:48 to 15:48 UT in steps of 12 min. The background images are from top to bottom $B_z$, H$\alpha$ and \ion{Ca}{ii} IR continuum intensity, H$\alpha$ line-core intensity, H$\alpha$ LOS velocity  and \ion{Ca}{ii} IR LOS velocity. The white rectangle in the rightmost column indicates the location of the ``perfect" IEF channel shown in Figure \ref{f:perfectief} below.  The red and blue arrows in the fifth column mark H$\alpha$ filaments.}\label{f:selected}
\end{figure*}
\subsubsection{Closed field lines in and near the sunspot}
The automatic selection of MFLs relevant for the IEF was based on a 150$\times$150 pixels (54\,Mm$\times$ 54\,Mm) square centered on the sunspot as initial seed points (see the black square in the left middle panel of Figure \ref{f:nfff}) for the magnetogram at 14:48 UT. We derived the MFLs originating in that area with a simplified approach using the Interactive Data Language. We rejected all MFLs that were open, whose two-dimensional (2D) length in the horizontal plane (see Section \ref{mag_topology}) was shorter than 5.4\,Mm and whose apex height exceeded 7.5\,Mm. That provided the spatial positions of 3915 suitable seed points for the VAPOR software \citep{atmos10090488} that we then used for all six time steps. The ad-hoc limits on the length and height in this step were based on the visibility and appearance of the IEF channels in the \ion{Ca}{ii} IR and H$\alpha$ lines, i.e., the spatial extent of fibrils and the formation heights of the lines.

From the VAPOR output, we then discarded all MFLs where the outer FP was located outside the de-projected aperture stop of IBIS and whose maximal height exceeded 7.5\,Mm, but did not require the minimal 2D length anymore. Table \ref{tab_select} lists the numbers and percentages of the MFLs remaining sequentially with each criterion for the magnetogram at  14:48 UT. This left on average 3158$\pm 74$ closed MFLs for one time step with a total of 18948 closed MFLs. This automatically selected sample will be labeled the ``large sample" in the following.  This sample was selected without considering the chromospheric velocities or intensities in the selection process in any way.

Figure \ref{f:full} shows 1/33 of the closed MFLs of the first and last magnetogram that were selected by this approach on top of the vertical component of the magnetic field $B_z$, the H$\alpha$ continuum intensity and the H$\alpha$ LOS velocity. The majority of the closed MFLs extends to the limb side of the sunspot from south to north with a cluster towards the north-east, where the following plage was located, because of the lack of opposite-polarity patches towards the west. MFLs from the umbra and inner penumbra were either open or violated the other two criteria listed above.

\subsubsection{Manually selected IEF channels}
The large sample of MFLs was complemented by a smaller sample of manually selected IEF channels. For that, we used the line-core intensity and velocity maps in both H$\alpha$ and \ion{Ca}{ii} IR to define about 15--25 seed points in each of the six maps that were located at the inner end of IEF channels for a total of 123 points. The points were chosen to be at about the end of flow channels or intensity fibrils and about evenly spaced around the sunspot in azimuth. Figure \ref{f:selected} shows these manually selected IEF channels for all six maps, with the MFLs resulting from the seed points. The manually selected sample is a subset of the large sample specifically chosen to catch flow channels and intensity fibrils. The MFLs towards the west are short and connect to moving magnetic features of opposite polarity close to the sunspot, while most MFLs towards the east do not align with the large and dark intensity filaments marked with arrows in Figure \ref{f:selected} that stand out prominently, especially in the \ion{Ca}{ii} IR line-core images.

 The two approaches of automatic and manual selection sample somewhat different targets. The automatic sample represents the magnetic topology of closed MFLs in the canopy of the sunspot with a good statistics regardless of the presence of IEF channels. It defines the characteristic topology in which IEF channels are found. The manual selection explicitly targets IEF, i.e., flow channels with a clear velocity signature, but at a worse statistics. Prior studies showed a rather isotropic distribution of IEF channels around sunspots, without a clear difference of IEFs to their surroundings in the photospheric magnetic field near their inner end points \citep{beck+choudhary2020}. It is not immediately obvious whether the latter is still valid when following the MFLs away from the sunspot and to chromospheric heights. The comparison of the two samples thus can be used to cross-check if there are special conditions for IEF channels to happen by any eventual differences in their average properties.

\begin{figure}
\resizebox{8.8cm}{!}{\includegraphics{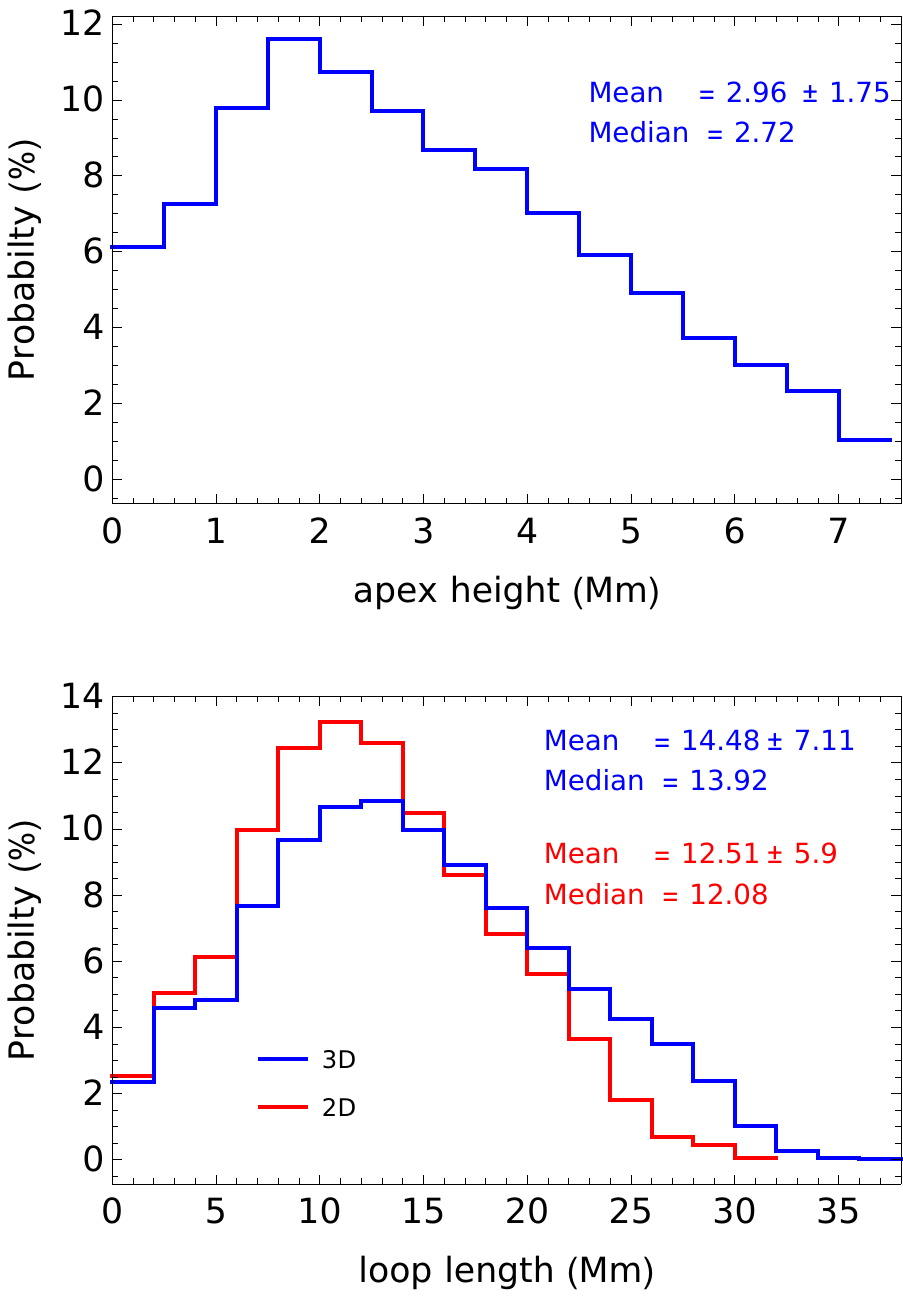}}
\caption{Histograms of the apex height (top panel) and the 3D (blue line) and 2D (red line) loop length (bottom panel) of closed MFLs in the large sample.}
\label{f:hist}
\end{figure}
\section{Results}\label{secres}
\subsection{Magnetic connectivity throughout the active region}
The top row of Figure \ref{f:nfff} shows the MFLs throughout the AR at 14:48 UT and  15:48 UT. The majority of closed MFLs connect from the sunspot to the trailing plage region. The MFLs west of the sunspot and east of the plage are primarily open and reach coronal heights. This part of the large-scale topology is reflected by the bright loops that are visible in the AIA 171\,{\AA} image in the bottom right panel of Figure \ref{f:nfff}. The large filament extending from the sunspot to the east can be traced in the AIA 171 and 304\,{\AA} images, but the no MFL passes along its axis. The MFLs near the filament instead form an arcade of loops crossing the polarity inversion line about perpendicular to the filament's spine. The closed MFLs originating inside or in proximity to the sunspot have an approximately radial orientation away from the sunspot center (Figures \ref{f:full} and \ref{f:selected}).

The HMI magnetograms showed little to no evolution within the 1-hr window of the observations (Figures \ref{f:nfff} and \ref{f:full}), hence the magnetic field extrapolation did not change significantly. The intensity and velocity maps in Figure \ref{f:selected} show some variation with time, e.g., the intensity pattern to the west changes completely in the H$\alpha$ line-core images, but all dark filaments to the east, north and south-west and most flow channels persist (see also the temporal average of the same time series in Figure 12 of \citeauthor{beck+choudhary2020} \citeyear{beck+choudhary2020}).

\begin{figure}
\resizebox{8.8cm}{!}{\includegraphics{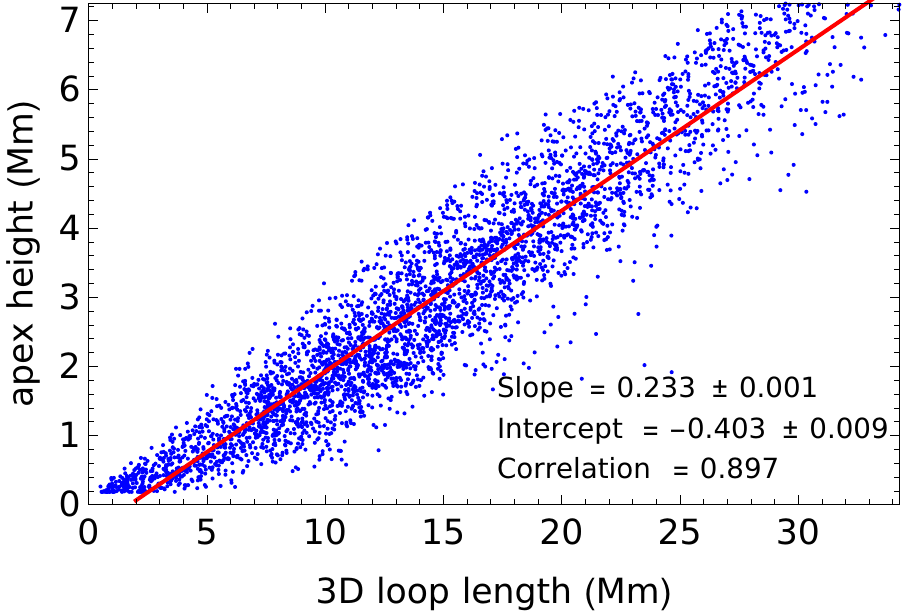}}
\caption{Scatter plot of 3D loop length and apex height for closed MFLs in the large sample.}
\label{f:scatterlength}
\end{figure}

\begin{figure}
\resizebox{8.8cm}{!}{\includegraphics{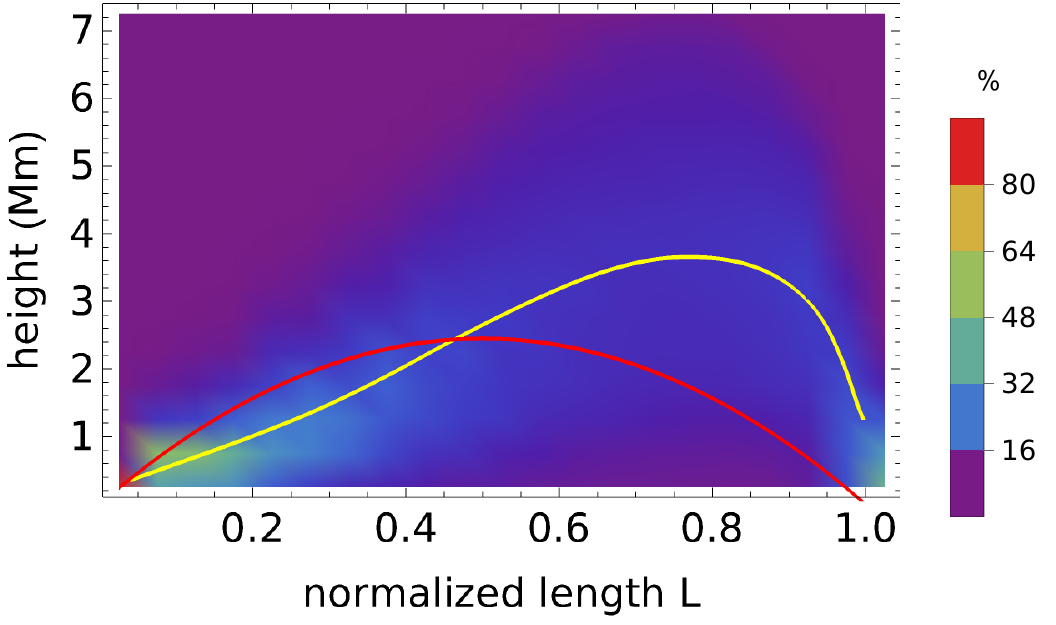}}
\caption{Loop shape for closed MFLs in the large sample against relative 2D length $L$. The background image shows the probability distribution of the height with the color bar at right. The inner (outer) FP is at $L=0$ (1). The yellow line indicates the average, while the red line is a parabolic loop with an apex height of 2.45\,Mm.}
\label{f:shape}
\end{figure}
\subsection{Properties of closed magnetic field lines in the large sample}
\subsubsection{Topology of closed MFLs: length, apex height, shape, $r_{\rm inner}, r_{\rm outer}$}\label{mag_topology}
For all closed MFLs, we defined the 3D loop length as the path length along the MFL between its two photospheric FPs, while the 2D loop length was defined as the length of their projection onto the horizontal plane. The apex height was defined as the maximum height attained along the MFL. The distances of the inner and outer FP from the sunspot center was measured both in absolute units and fractional to the outer penumbral radius of $r_0 = 16.3$\,Mm, where the inner FP was defined as the seed point of the MFL inside the square centered on the sunspot.

The top panel of Figure \ref{f:hist} shows the histogram of the apex height of all closed MFLs in the large sample. The distribution is roughly Gaussian between 0 and 7.5\,Mm with a steeper drop off towards larger heights. The average and median values are 2.96\,Mm and 2.72\,Mm, respectively, far from the rejection threshold of 7.5\,Mm height. Those heights should well fall into the formation height range of the H$\alpha$ line, but should exceed the one of \ion{Ca}{ii} IR. The histograms of the 2D and 3D loop length are fairly similar (bottom panel of Figure \ref{f:hist}) ranging from 0 to about 30\,Mm length with average values of about 12--14$\pm 6$\,\,Mm.

There is a close relation between the 3D loop length and the apex height (Figure \ref{f:scatterlength}), which holds in the same way for the 2D loop length (not shown). The correlation coefficient between length and height is about 0.9 with a ratio $L:H$ of about 4:1.

\begin{figure}
\resizebox{8.8cm}{!}{\includegraphics{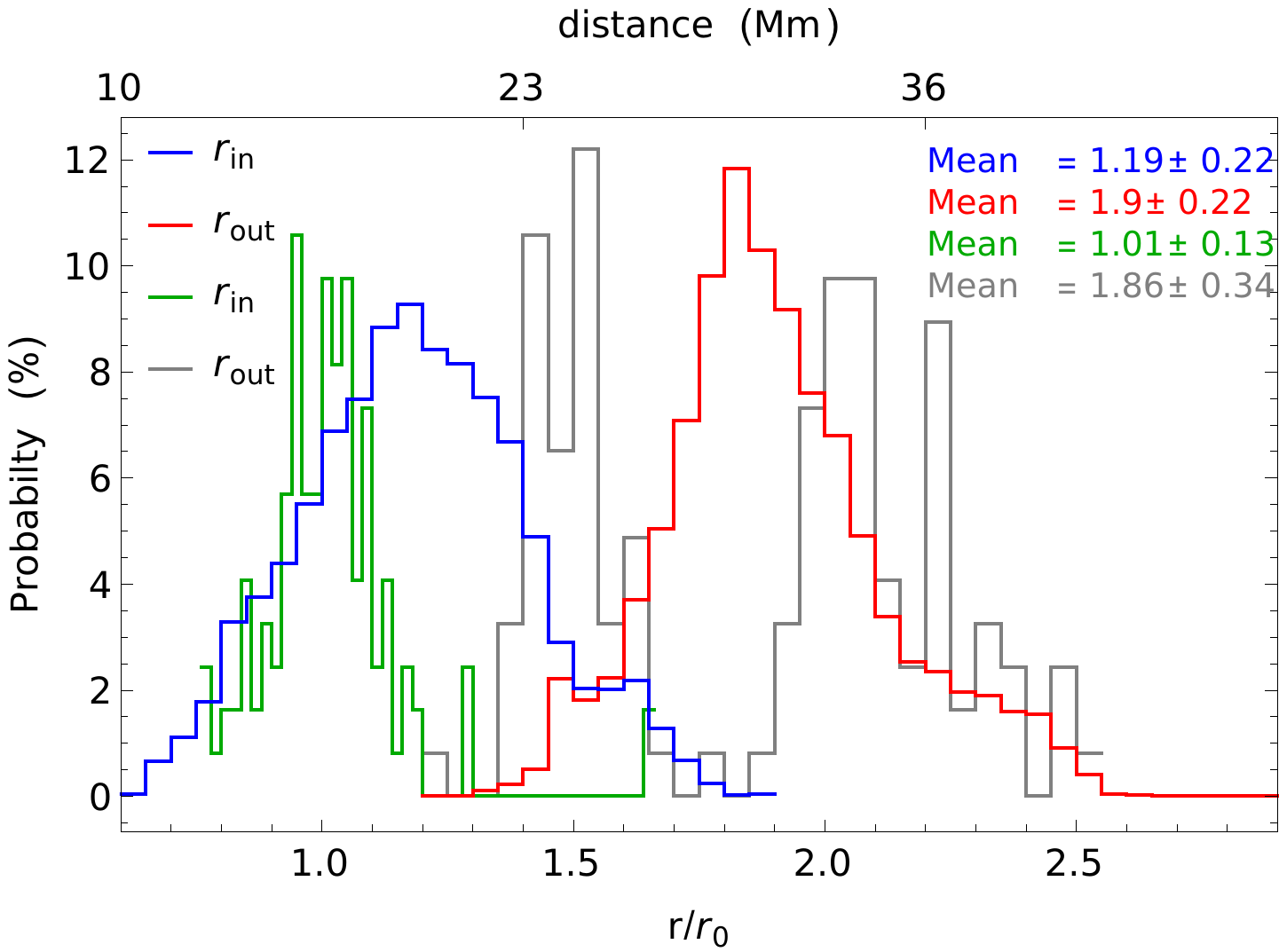}}
\caption{Histograms of the radial distance of the inner (blue line) and outer FPs (red line) of closed MFLs from the center of the sunspot in the large sample.  The green and gray lines show the corresponding histograms for the manually selected sample. The x-axis at the top (bottom) gives the values in Mm (fractional  outer penumbral radius  $r_0$).}
\label{f:distance}
\end{figure}

\begin{figure}
\resizebox{8.8cm}{!}{\includegraphics{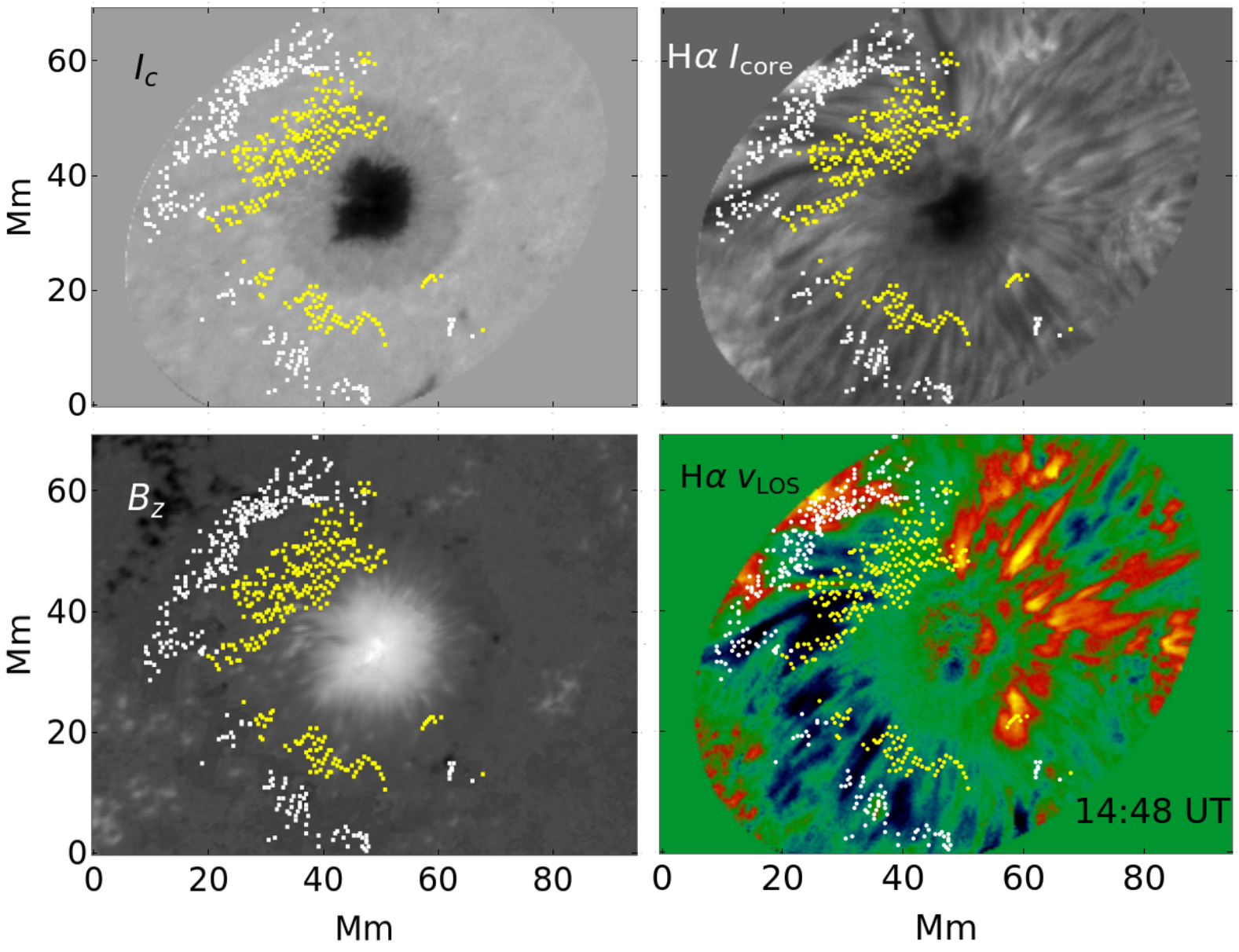}}
\caption{Locations of the inner and outer FPs of closed MFLs at 14:48 UT. The images in the background show (clockwise, starting left top) the H$\alpha$ continuum and line-core intensity, the H$\alpha$ LOS velocity, and $B_z$. Yellow (white) points denote the locations of the inner (outer) FPs.}
\label{f:footpoints}
\end{figure}

To determine the typical loop shape, we first normalized the 2D length $L$ of each closed MFL to be from 0 to 1 by dividing it with the full length of the MFL. We then calculated the histograms of the height in 20 length bins of 0.05 extent each and sorted the result into a $L$-height array. Figure \ref{f:shape} shows a slightly smoothed 2D representation of these histograms of the height at each relative length point together with its average value. The closed MFLs form arched loops with a larger inclination to the vertical at the inner FPs that are nearly vertical at the outer FPs. The average apex height of 3.65\,Mm is attained at a relative length of 0.8 close to the outer FP. Heights below 1\,Mm and above 6\,Mm are rarely attained ($< 16\%$) at any place. The shape roughly matches the parabolic loop of 2.45\,Mm height  (red line in Figure \ref{f:shape}) that was inferred as best match to the velocity maps of this sunspot in \citet{beck+etal2020}.

\begin{figure} 
\resizebox{8.8cm}{!}{\includegraphics{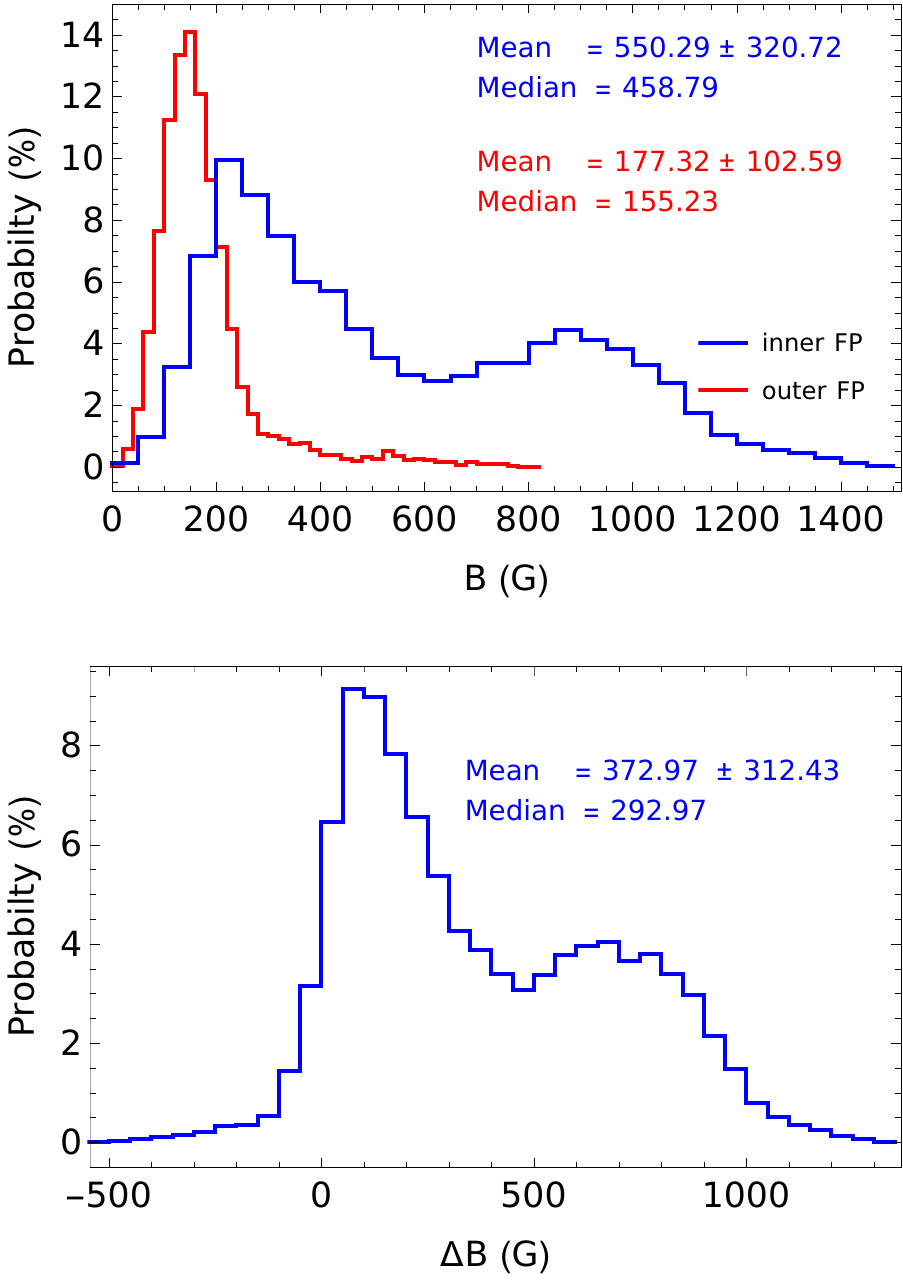}}
\caption{Histograms of the magnetic field strength $B$ (top panel) at the inner (blue line) and outer (red line) FPs, and of their difference $\Delta B = B_{\rm inner} - B_{\rm outer}$ (bottom panel).}
\label{f:histblarge}
\end{figure}

\begin{figure}
\resizebox{8.8cm}{!}{\includegraphics{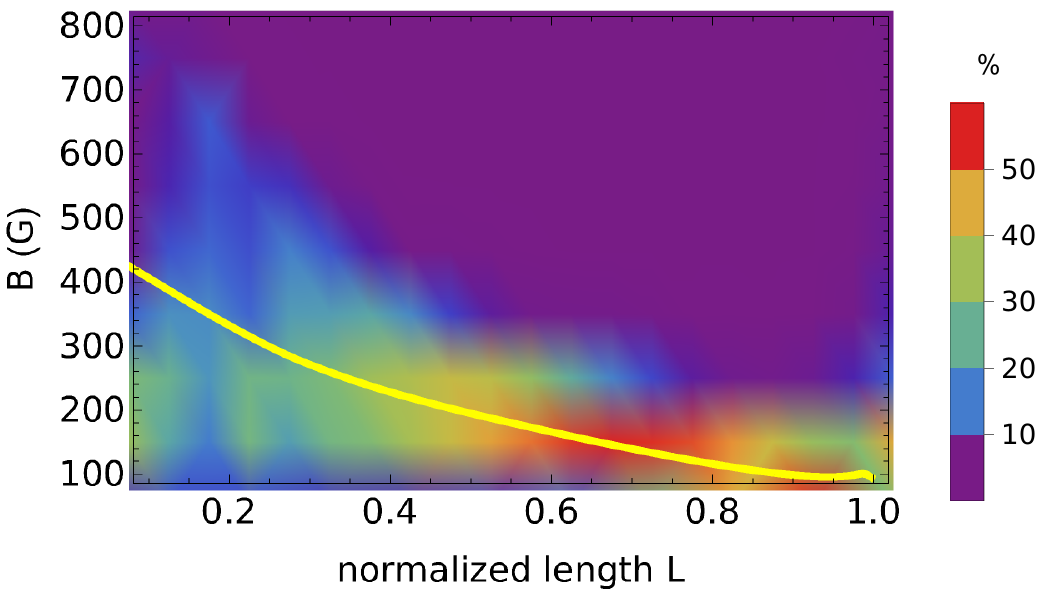}}
\caption{Field strength for closed MFLs in the large sample against relative 2D length $L$. The background image shows the probability distribution of $B$ with the color bar at right. The inner (outer) FP is at $L=0$ (1). The yellow line indicates the average value.}
\label{f:bshape}
\end{figure}

\begin{figure*}
\resizebox{16cm}{!}{\includegraphics{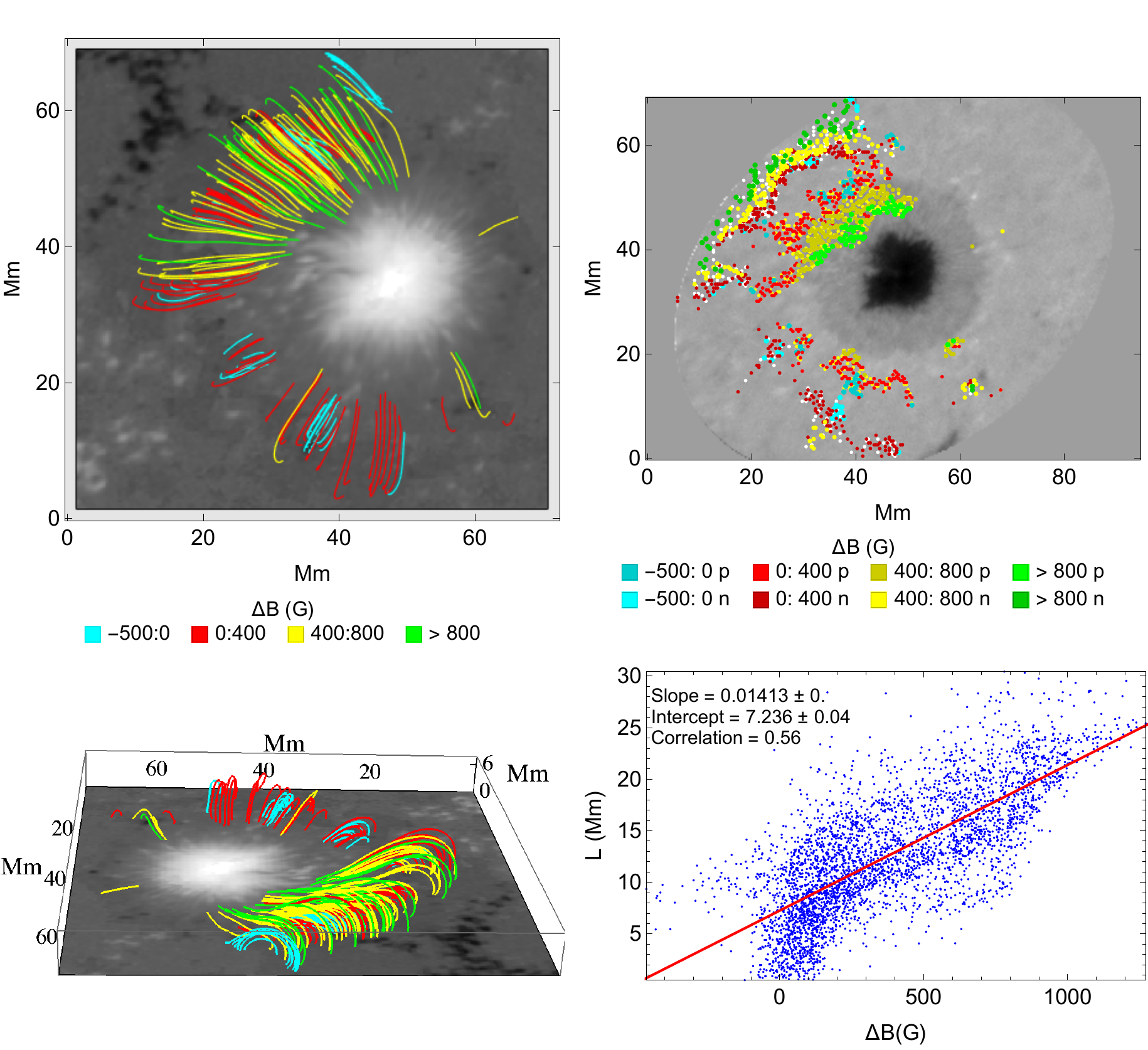}}
\caption{Properties of MFLs with different $\Delta B$. Top left: field lines with $\Delta B <0$\,G (turquoise), 0--400\,G (red), 400--800\,G (yellow), and $> 800$\,G (green) in a top view overlaid on $B_z$. Bottom left: the same in a side view to highlight the shape. Top right: inner and outer FPs for the same ranges in $\Delta B$. Bottom right: scatter plot of $\Delta B$ and 3D loop length. The map in the bottom left panel is rotated relative to the images in the top row for a better visibility of the closed MFLs.} \label{f:neg}
\end{figure*}
Figure \ref{f:distance} shows the histograms of the distance of the inner and outer FPs from the center of the sunspot, while Figure \ref{f:footpoints} shows their location on top of maps of photospheric and chromospheric quantities at 14:48 UT. The inner FPs can be found at a radial distance of 0.6--1.8\,$r_0$ with an average value of 1.19\,$r_0$, slightly outside the outer penumbral boundary. None are found inside the umbra, where the field lines were open. The outer FPs are at 1.5--2.5\,$r_0$ with an average distance of 1.9\,$r_0$ in primarily plage regions of opposite polarity (Figure \ref{f:footpoints}). Only the short closed MFLs to the south-east ($x,y = 20,20)$ and south-west ($x,y = 60,20)$ connect to magnetic elements in the sunspot moat. The locations of the inner FPs coincide to a large extent with the end points of flow fibrils (bottom right panel of Figure \ref{f:footpoints}).

\subsubsection{Magnetic field strength}
Figure \ref{f:histblarge} shows the histograms of the magnetic field strength $B$ from the HMI Milne-Eddington inversion at the inner and outer FPs of closed MFLs. The magnetic field strength at the inner FPs ranges between 0--1400\,G with an average value of 550\,G. The histogram shows a double-peaked distribution with peaks at 250 and 900\,G, where the latter peak and the extension to 1400\,G corresponds to inner FPs inside the penumbra. The histogram for the outer FPs has a roughly Gaussian distribution from 0 to 400\,G with an average value of 180\,G. The difference in field strength between inner and outer FPs $\Delta B$ (bottom panel of Figure \ref{f:histblarge}) ranges from -500 to +1300\,G with an average value of +370\,G and a similar bi-modal shape as for the inner FPs. About 6\,\% of the MFLs show a negative value of $\Delta B$ with higher field strength at the outer FPs.

Figure \ref{f:bshape} shows how the field strength along the MFLs varies between the FPs. The field strength $B(x,y,z)$ was determined at the corresponding height of the MFL at each spatial position $(x,y)$ in the horizontal plane, and then plotted against the 2D length $L$. It drops monotonically from the average value of 450\,G with a broad distribution at the inner end up to a relative length of about 0.95. At that length, almost all the MFLs have a similar value of $B$ of 180\,G that represents a small local maximum in $B$. At the inner and outer FPs, the MFLs are close to the photosphere, while they sample chromospheric layers in between (Figure \ref{f:shape}).

Figure \ref{f:neg} identifies the locations of inner and outer FPs in four bins in $\Delta B$ of 400\,G width each for the data at 14:48 UT. Most MFLs with $\Delta B<0$\,G belong to comparably short loops, whose inner FPs are in some cases far outside the outer penumbral boundary. For MFLs with increasing values of $\Delta B$ the inner FPs move towards the umbra and the loop height increases, while the outer FP locations do not vary much. The MFLs with $\Delta B >800$\,G are found on top of all others (bottom left panel of Figure \ref{f:neg}). The scatter plot in the bottom right panel of Figure \ref{f:neg} shows a high correlation between the 3D loop length and $\Delta B$, in line with the displacement of the inner FPs towards the umbra for increasing $\Delta B$, which also increases the length.

\begin{figure}
\resizebox{8.8cm}{!}{\includegraphics{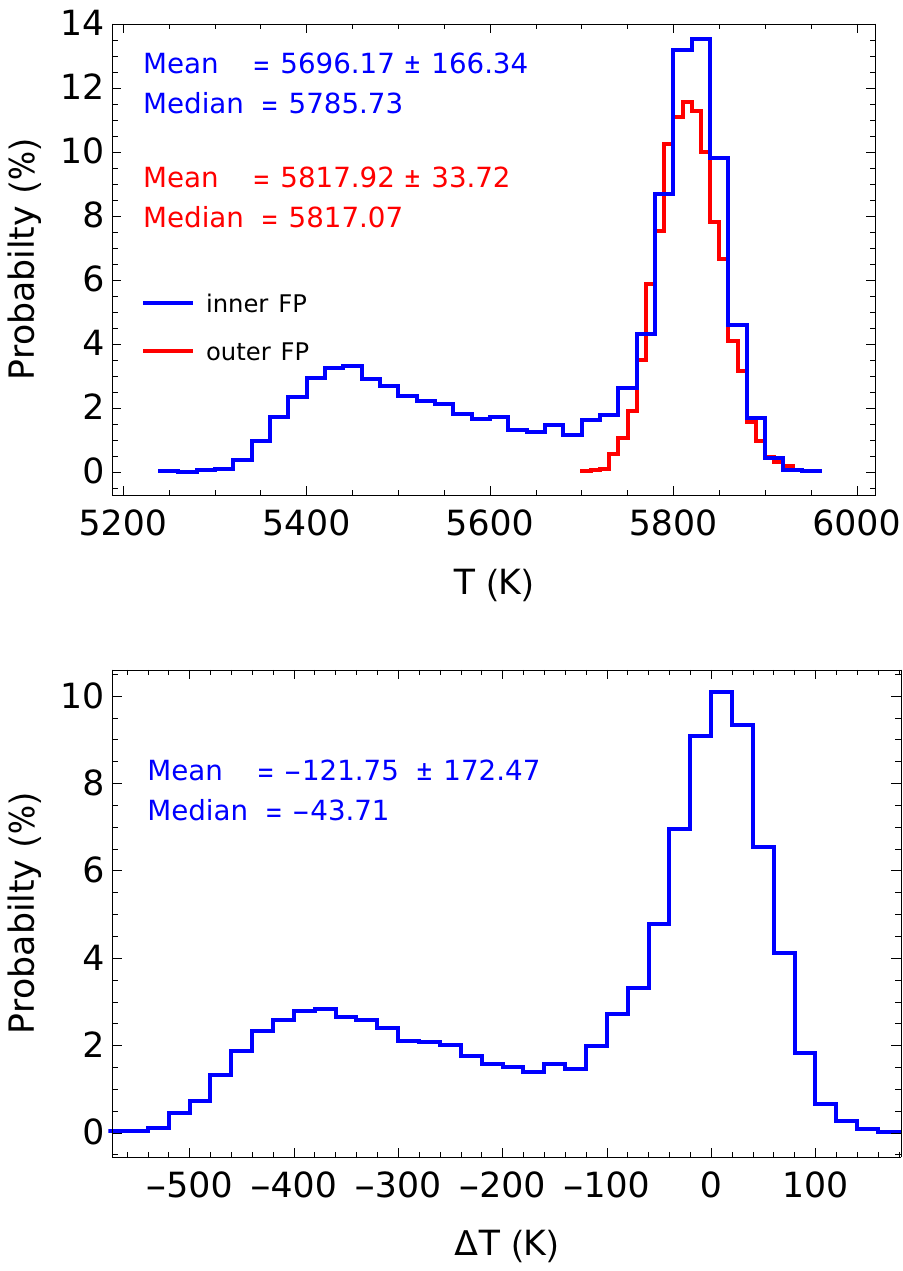}}
\caption{Histograms of the temperature $T$  (top  panel)  at  the  inner  (blue  line) and  outer  (red  line)  FPs,  and  of  their  difference $\Delta T = T_{\rm inner} - T_{\rm outer}$ (bottom panel).}\label{f:temperature}
\end{figure}

\subsubsection{Continuum temperature}
Figure \ref{f:temperature} shows the histograms of the temperatures at the inner and outer FPs, and their difference $\Delta T = T_{\rm inner} - T_{\rm outer}$. Similar to the magnetic field strength, the inner FPs exhibit a bi-modal distribution with one peak at the quiet sun (QS) temperature of about 5800\,K, an extended tail to lower temperatures down to 5300\,K, and a smaller peak at 5450\,K corresponding to FPs inside the penumbra. The distribution for the outer FPs is Gaussian-shaped around the QS temperature with an average value of $5820\pm 34$\,K. The temperature difference $\Delta T$ is primarily negative from -500 to +100\,K with two peaks at -350 and 0\,K, again caused by the location of the inner FPs in the QS or the cooler penumbra. About 32\,\% of the closed MFLs show a positive temperature difference, which could drive a flow opposite to the IEF.

\begin{figure}
\resizebox{8.8cm}{!}{\includegraphics{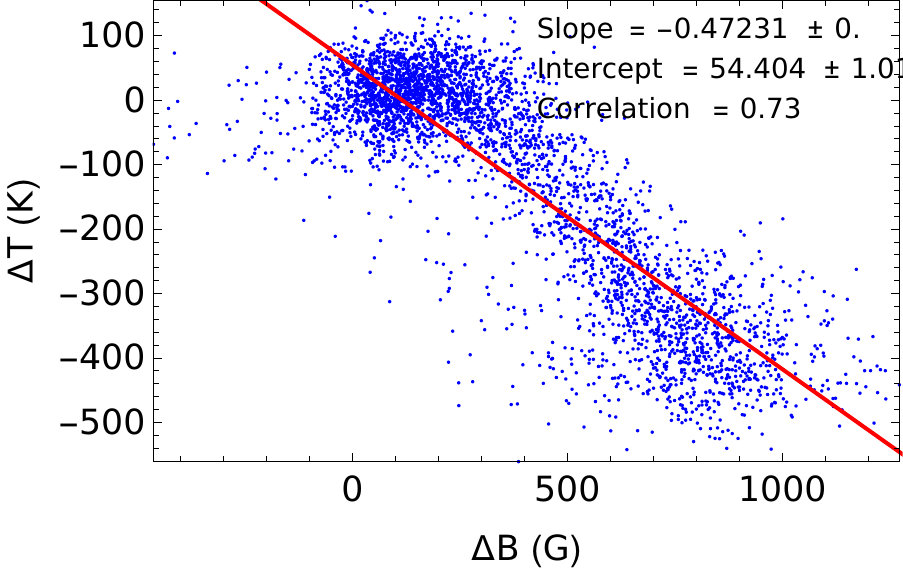}}
\caption{Scatter plot of the difference between inner and outer FPs in temperature $\Delta T$ and field strength $\Delta B$. }
\label{f:temp_b}
\end{figure}

The high correlation between $\Delta T$ and $\Delta B$ of 0.73 in their scatter plot in Figure \ref{f:temp_b} confirms a common origin for the behavior of the temperature and field strength difference, i.e., the radial gradient in both $B$ and $T$ with distance from the sunspot center and the displacement of inner FPs with higher $\Delta B$ towards the umbra. Both $\Delta B >0$ and $\Delta T <0$ create a gas pressure difference in the same direction along a closed MFL, so their anti-correlation amplifies the potential driving force.

\begin{figure*}
\resizebox{17cm}{!}{\includegraphics{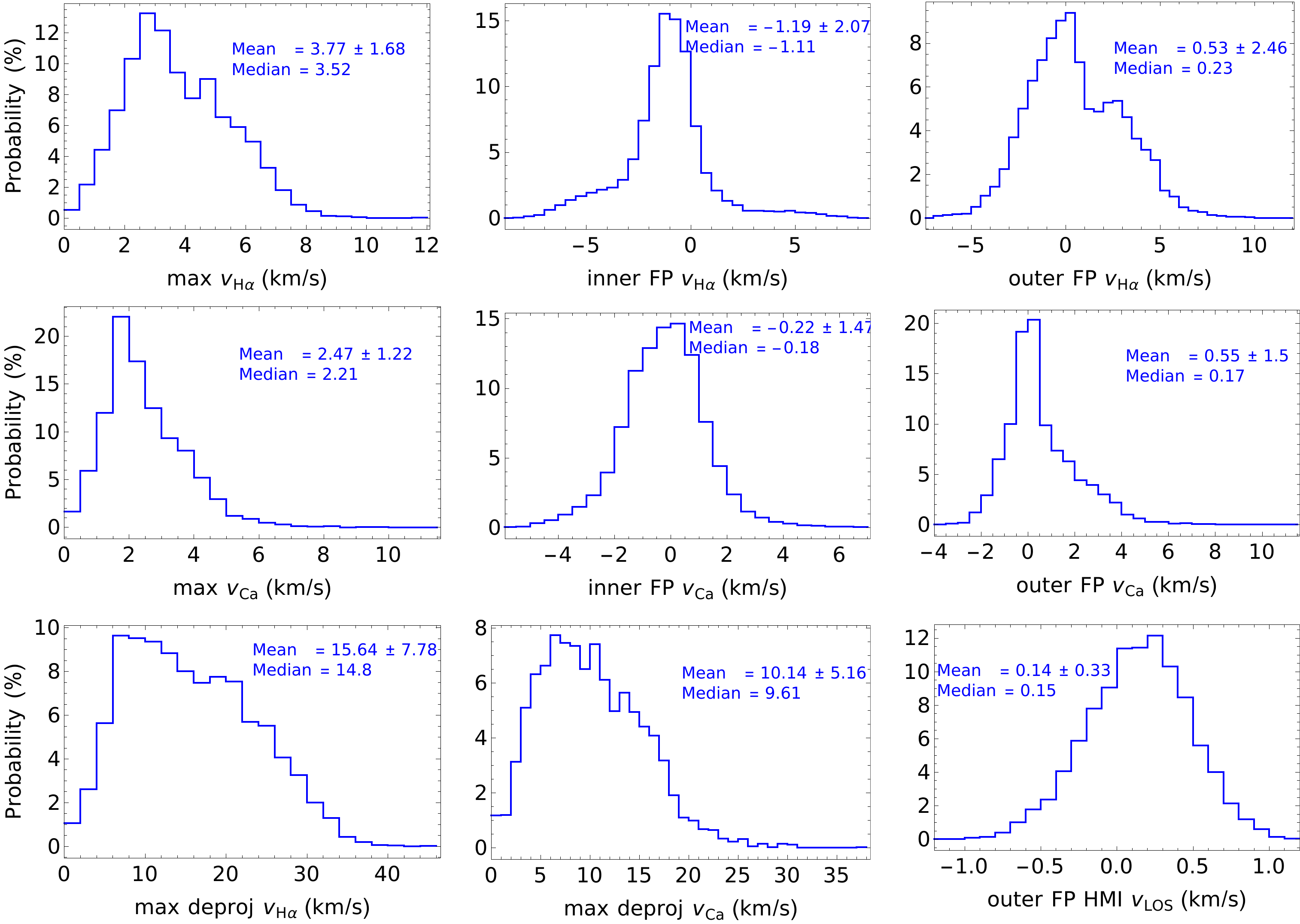}}
\caption{Histograms of chromospheric and photospheric velocities. Top row, left to right: maximal H$\alpha$ LOS velocity along field lines, and H$\alpha$ LOS velocities at the inner and outer FPs. Middle row: the same for \ion{Ca}{ii} IR. Bottom row, left to right: de-projected maximal velocities in H$\alpha$ and \ion{Ca}{ii} IR, and photospheric LOS velocity from HMI at the outer FPs.}\label{f:vel_hist}
\end{figure*}

\begin{figure*}
\resizebox{17cm}{!}{\includegraphics{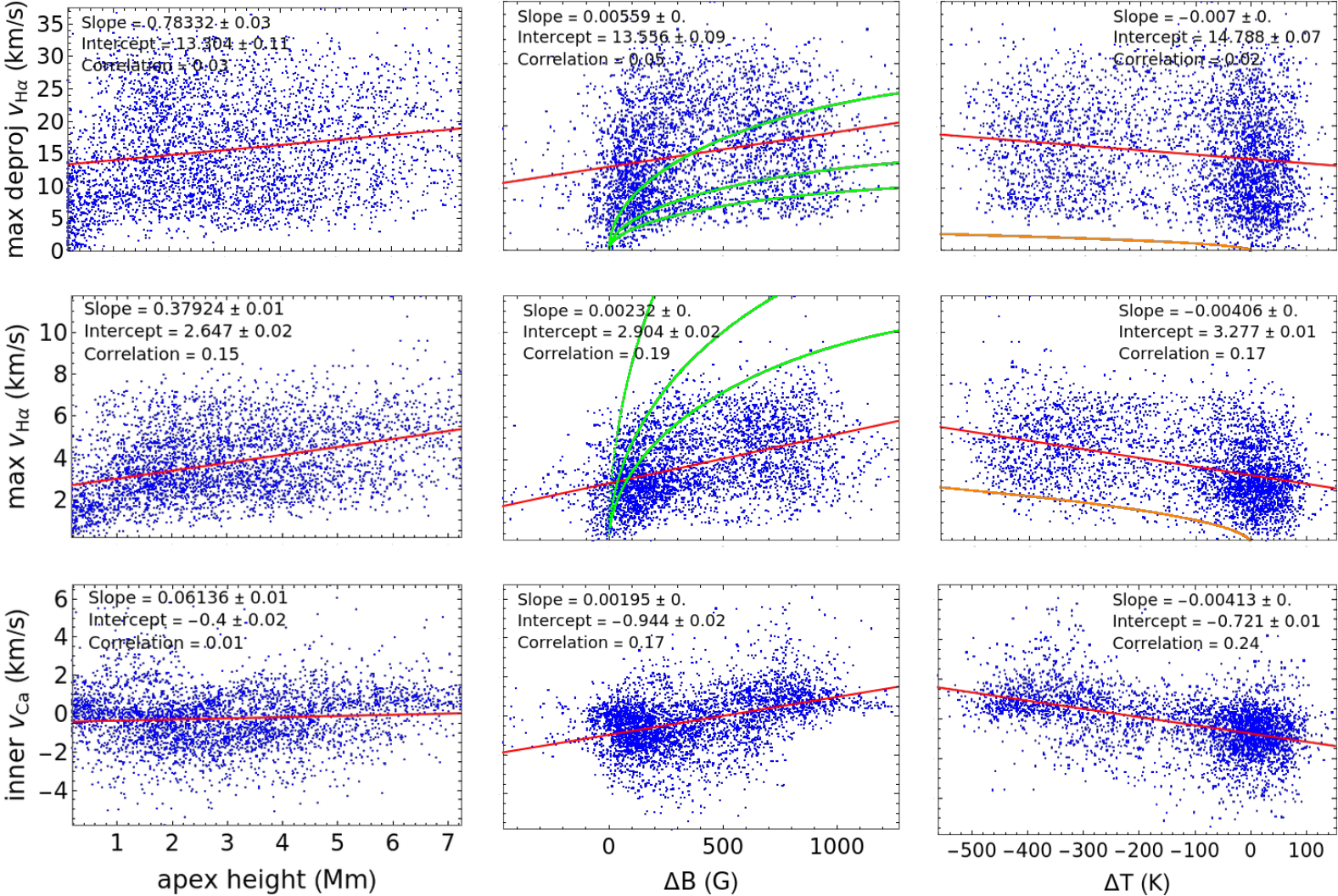}}
\caption{Scatter plots of apex height (left column), $\Delta B$ (middle column) and $\Delta T$ (right column) against flow velocities. Top to bottom: maximal de-projected H$\alpha$ velocity, maximal H$\alpha$ LOS velocity, and \ion{Ca}{ii} IR LOS velocity at the inner FP. The green lines in the middle column show the predicted velocity from the field strength difference for HSRA gas densities at (the lowest line to highest) $\log \tau = -0.3, -1$ and $-2$. The orange lines in the right column show the flow speed predicted by the temperature difference.} \label{f:vel_scat}
\end{figure*}

\begin{figure*}
\resizebox{17cm}{!}{\includegraphics{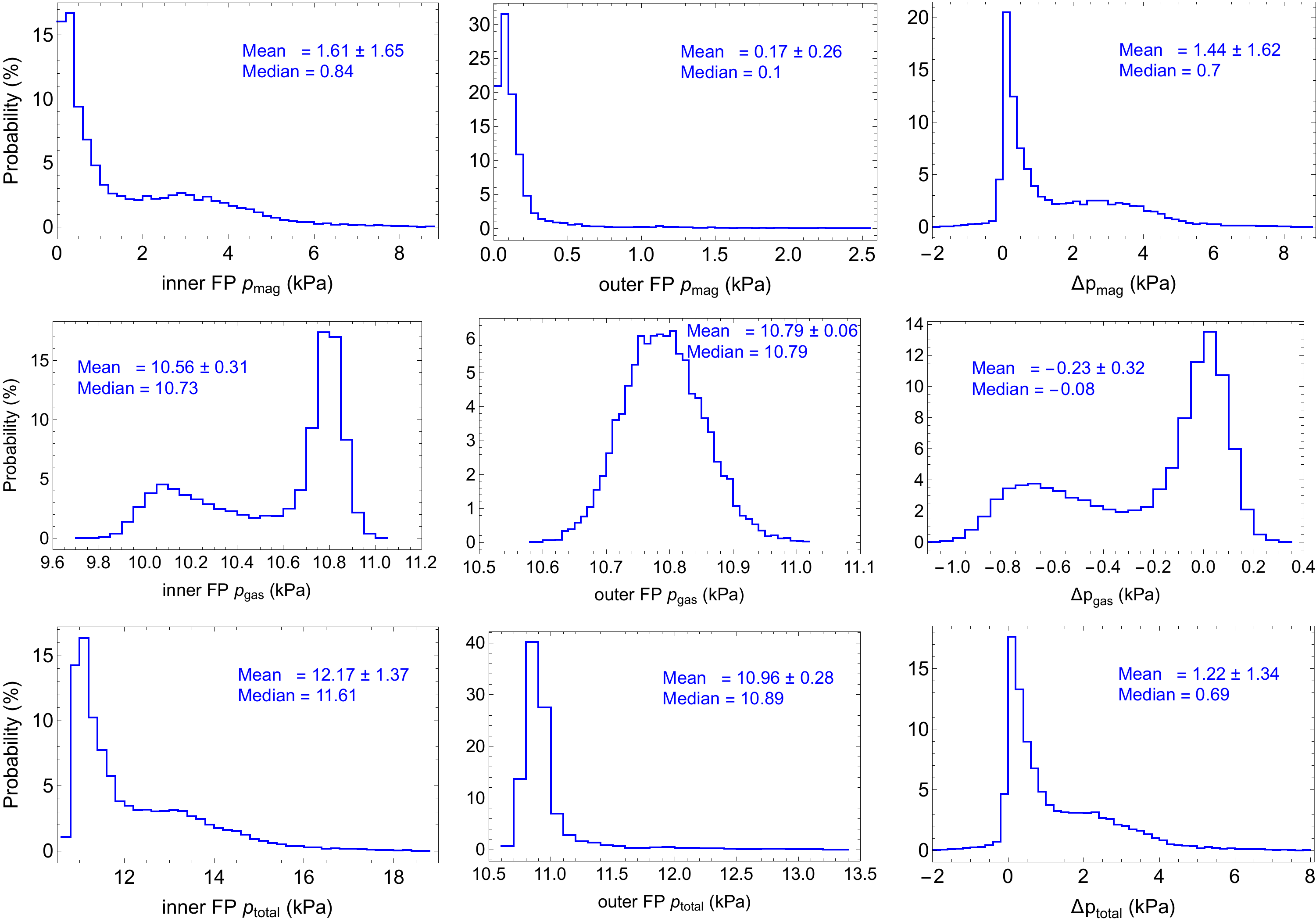}}
\caption{Histograms of the magnetic pressure (top row), gas pressure (middle row) and total pressure (bottom row). Left to right: pressures at the inner and outer FPs, and their difference $\Delta p = p_{\rm inner} - p_{\rm outer}$.}\label{f:pressurehist}
\end{figure*}
\subsubsection{Photospheric and chromospheric velocities}
For the two chromospheric spectral lines of H$\alpha$ and \ion{Ca}{ii} IR, we determined the velocity at the inner and outer FPs and the maximal unsigned velocity along each closed MFL. The latter was defined as the maximum velocity encountered along the 2D paths of the MFLs in the horizontal plane, ignoring the height of the loop and any projection effects. The maximal velocity was derived from both the LOS and de-projected velocity maps. We only retrieved the velocity at the outer FPs from the photospheric HMI velocities to check for possible upflows at the outer end, as the velocities at the inner FPs are expected to be strongly contaminated with the regular Evershed flow.

The top two rows of Figure \ref{f:vel_hist} show the histograms of the three velocities ($v_{\rm max}$, $v$ at inner and outer FPs) as derived from the LOS velocities of both chromospheric lines. The maximal flow speeds along closed MFLs range from 0 to 8--10\,km\,s$^{-1}$ with averages of 2.5 and 3.8\,km\,s$^{-1}$ for \ion{Ca}{ii} IR and H$\alpha$, respectively. All chromospheric velocities at the inner and outer FPs scatter around roughly zero with averages below $\pm 0.6$\,km\,s$^{-1}$ apart from the velocity at the inner FPs in H$\alpha$ with an average of -1.2\,km\,s$^{-1}$, which presumably reflects the larger area coverage of the inner FPs on the limb side with its large-scale blue shift pattern. The maximal velocities in the de-projected velocity maps (bottom left two panels in Figure \ref{f:vel_hist}) largely mirror the same quantity in the LOS velocities, just with an extended range from 0--30\,km\,s$^{-1}$ and average values of 10--15\,km\,s$^{-1}$, i.e., an increase by a factor of about 5. The photospheric velocity at the outer FPs shows a slight red shift of $0.14\pm 0.33$\,km\,s$^{-1}$ on average, but with a large scatter from -1 to +1\,km\,s$^{-1}$. Even with a correction for the convective blue shift of -0.2\,km\,s$^{-1}$ (see Appendix \ref{app_photvelo}), no systematic upflows are seen in the photosphere at the outer FPs.

Figure \ref{f:vel_scat} shows scatter plots of the LOS and de-projected velocities against the apex height, $\Delta B$ and $\Delta T$. None of the velocities exhibits a strong correlation with the apex height. The majority of the data points clusters around zero on the abscissa in all plots of velocities against $\Delta B$ or $\Delta T$ (right two columns of Figure \ref{f:vel_scat}). The plots against $\Delta B$ clearly reveal that only a small fraction of MFLs have  $\Delta B <0$, while $\Delta T >0$ happens more frequently in comparison.

All plots of either LOS or de-projected maximal H$\alpha$ velocities have upper and lower boundary regions of minimal or maximal velocities that rarely occur. The lower boundary is better defined and indicates the absence of small velocities below $<2$\,km\,s$^{-1}$ in the LOS and $<4$\,km\,s$^{-1}$ in the de-projected velocities for large values of $\Delta B > 500$\,G or $\Delta T < -200$\,K. The shape of the distributions matches to first order the predicted square-root dependence of the flow speed on the magnetic or gas pressure differences. Using Equation (\ref{speed_eq}), we calculated the expected velocities for a given value of $\Delta B$ and $\Delta T$ (solid green and orange lines in Figure \ref{f:vel_scat}). The latter is independent of the gas density, while for the former we used three different gas densities in the HSRA model at $\log \tau = -0.3, -1$ and $-2$. With the close anti-correlation of $\Delta B$ and $\Delta T$ (Figure \ref{f:temp_b}), the total driving force is expected to be larger than the individual separate contributions. Both the observed LOS and maximal H$\alpha$ velocities exceed a purely thermal driver (right column of Figure \ref{f:vel_scat}). In case of a driver based on the difference in magnetic field strength, the predicted velocities in the middle column of Figure \ref{f:vel_scat} exceed the observed maximal H$\alpha$ LOS velocities, but fall into the observed range for the de-projected velocities. Depending on the height in the solar chromosphere, the sound speed as the upper limit of mass flows is 6--20\,km\,s$^{-1}$. The highest, but still rather low correlation values of about 0.2 are found for the relation between $\Delta B$ or $\Delta T$ and the velocities in \ion{Ca}{ii} IR at the inner FPs (bottom right two panels of Figure \ref{f:vel_scat}), where the downflow points of the IEF channels would be expected.

\subsubsection{Pressure balance}
Figure \ref{f:pressurehist} shows the histograms of the magnetic, gas and total pressure at the inner and outer FPs, and their differences $\Delta p = p_{\rm inner} - p_{\rm outer}$. For all calculations that include the gas pressure, the photospheric density in the HSRA at $\log \tau = -0.2$ of 2.89$\times 10^{-7}$\,g\,cm$^{-3}$ was used.

The magnetic pressure at the inner FPs ranges from 0 to 8\,kPa s with an average of 1.6\,kPa. The corresponding values at the outer FPs are much smaller, with 0--0.5\,kPa and an average of 0.17\,kPa. The resulting difference $\Delta p_{\rm mag}$ is therefore mainly positive, with a similar total range as for the inner FPs. In the same way as for temperature, the gas pressure at the inner FPs has a bi-modal distribution with one peak at about 10.1\,kPa corresponding to the penumbra and one at about 10.8\,kPa corresponding to the QS, while the histogram for the outer FPs only shows the latter. The gas pressure difference is primarily negative, down to -1\,kPa. The total pressure (bottom row of Figure \ref{f:pressurehist}) at the inner FPs with values from 11 to 16\,kPa exceeds the one at the outer FPs that is fairly constant at 10.96 $\pm 0.3$\,kPa. This leads to a total pressure imbalance of up to +5\,kPa under the assumption of equal gas densities, while 9\,\% of the MFLs have a negative total pressure difference.

\begin{figure} 
\resizebox{8.8cm}{!}{\includegraphics{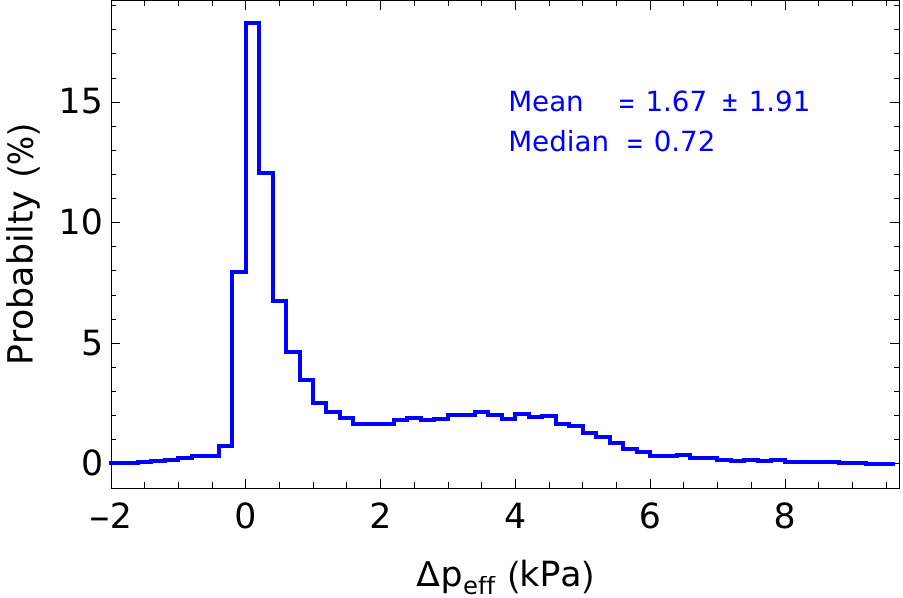}}
\caption{Histogram of the effective pressure difference $\Delta p_{\rm eff}$.}
\label{f:eff_pressure}
\end{figure}

\begin{table}
\caption{Line fit and physical parameters derived from the $T(B^2)$ relation.}\label{tab_densfit}
\begin{tabular}{c|cccc}
 & inner FPs & QS & penumbra & umbra\cr\hline\hline
slope & -0.0023 & -0.0014 & -0.0083 & - 0.0044\cr
intercept &13.5 & 8.5 & 46.3 & 23.4 \cr
correlation & 0.85 & 0.26 & 0.33 & 0.50 \cr
$\rho/10^{-7}$ (g\,cm$^{-3}$)& 2.69 & 4.31 & 0.75 & 1.42\cr
\hline
$\langle T \rangle$ (K) & 5696 & 5796 & 5417 & 3994\cr
$\langle B \rangle$ (kG) & 0.55 & 0.39 & 1.26 & 2.45\cr
$p_{\rm gas}$ (kPa) & 9.83 & 15.97 & 2.60 & 3.64\cr
$p_{\rm mag}$ (kPa) & 1.20 & 0.6 & 6.27 & 23.75\cr
$p_{\rm tot}$ (kPa) & 11.03 & 16.57 & 8.87 & 27.39\cr
\hline
\end{tabular}
\end{table}
\begin{figure*}
\resizebox{17cm}{!}{\includegraphics{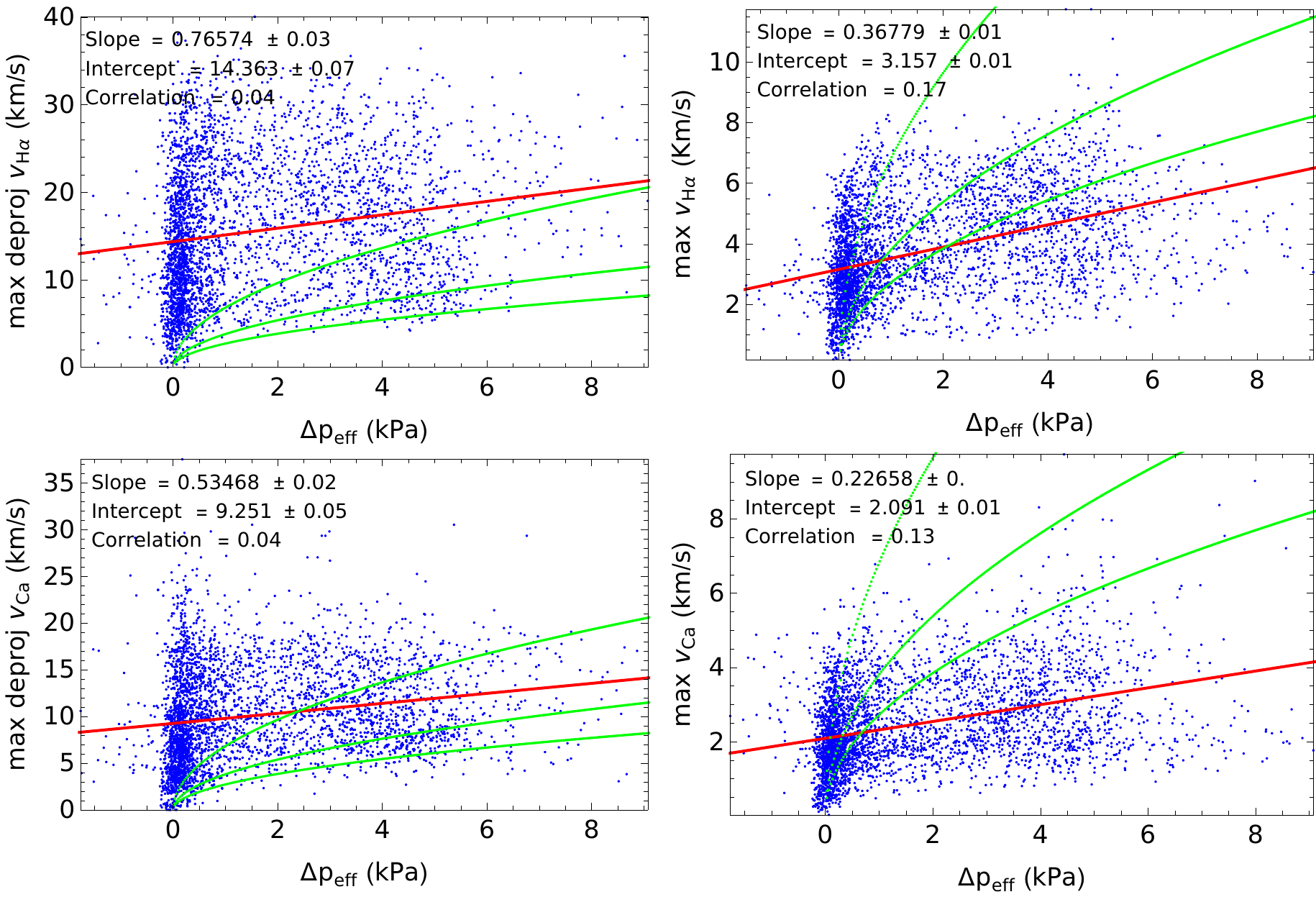}}
\caption{Scatter plots of the effective pressure difference $\Delta p_{\rm eff}$ against de-projected (left column) and LOS (right column) flow velocities. The green lines show the predicted velocities for HSRA gas densities at $\log \tau = -0.3, -1$ and $-2$.}\label{f:pressure_scat}
\end{figure*}

As we cannot confirm at once that the assumption of a global and large-scale total pressure balance holds over the comparably large distances between the inner and outer FPs of closed MFLs and the characteristic values of both $\Delta B$ and $\Delta T$ would drive a flow in the same direction, we calculated the "effective" pressure difference following Equation (\ref{p_effect}) in addition (Figure \ref{f:eff_pressure}). It yields slightly higher pressure differences from -2 to 8\,kPa, but otherwise follows the shape of the corresponding histograms of $\Delta B$ and $\Delta T$ with only a small fraction of MFLs with $\Delta p_{\rm eff} <0$.

\begin{figure*}
\resizebox{17cm}{!}{\includegraphics{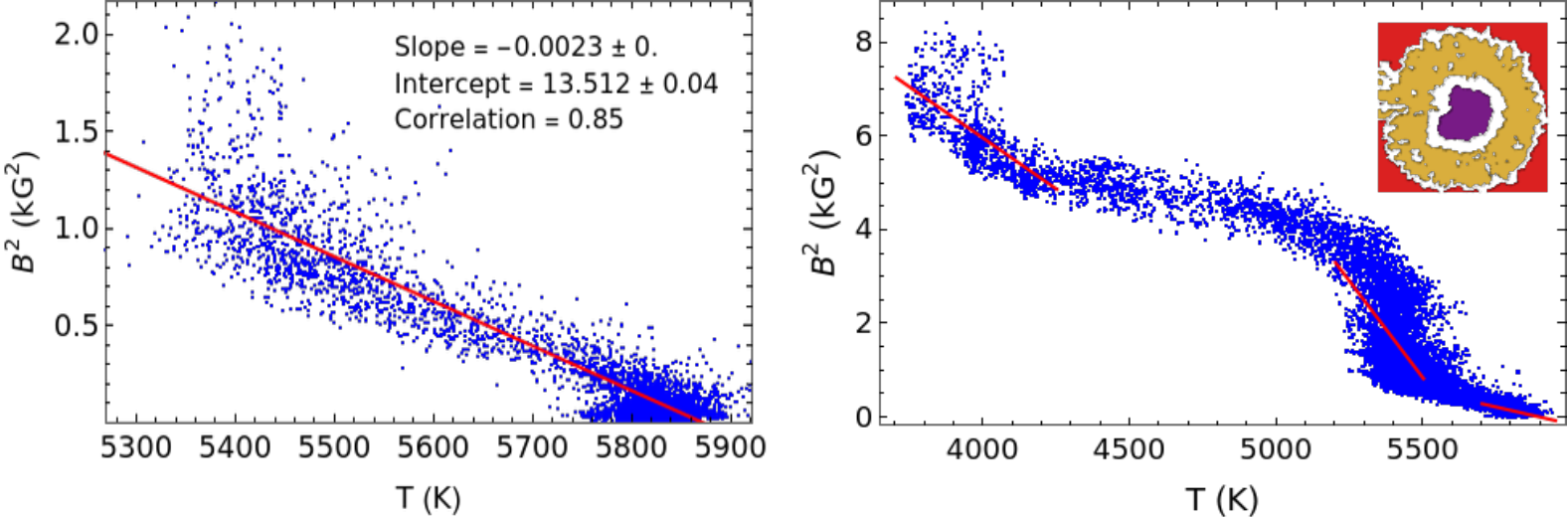}}
\caption{Scatter plots of the relation between $T$ and $B^2$. Left panel: for the inner FPs of the large sample. Right panel: for a square covering the whole sunspot and some QS regions in all six time steps. The red lines are straight line fits to all or subsets of data points. The inset in the upper right corner of the right panel shows the corresponding spatial mask for the umbra (purple), penumbra (orange) and QS (red) at  14:48 UT.}\label{f:tvsb}
\end{figure*}

The scatter plots of $\Delta p_{\rm eff}$ against maximal de-projected and LOS velocities in Figure \ref{f:pressure_scat} show a similar behavior as before for $\Delta B$ and $\Delta T$ alone (Figure \ref{f:vel_scat}). The shape of the square-root dependence of the predicted velocities with its lower boundary is matched. The observed LOS velocities fall short of the prediction, while the de-projected velocities span the whole range of predicted velocities when using three different gas densities. We only varied the gas density in the application of Equation (\ref{speed_eq}), but not in the calculation of $\Delta p_{\rm eff}$ that was derived with the HSRA gas density value at $\log \tau = -0.2$.

The question which gas density is appropriate across a 2D FOV on the solar surface is in general difficult to answer without additional assumptions. We thus tried to derive suitable gas densities using the relation between $T$ and $B^2$ as predicted by Equation (\ref{eq_density}) based on the photospheric HMI temperature and magnetic field strength. The left panel of Figure \ref{f:tvsb} shows the scatter plot of $T$ and $B^2$ for the inner FPs of the large sample, while the right panel shows the same for a square area covering all the umbra, penumbra and some QS areas outside the outer penumbral boundary. The inner FPs were located close to the outer penumbral boundary, primarily slightly outside the sunspot, with a temperature range of 5250--5950\,K. The relation between $T$ and $B^2$ for the inner FPs has a high correlation of about 0.85. The right panel reveals three different regimes in QS, penumbra and umbra, with a distinct change of behavior at the outer umbral and penumbral boundaries. We thus fitted three straight lines to the values in the right panel using a spatial mask for the three regimes (see the inset in the right panel of Figure \ref{f:tvsb}) excluding the areas of transition between them. Table \ref{tab_densfit} lists the resulting fit parameters and the derived physical quantities gas density $\rho$ and total pressure $p_{\rm tot}$. The correlation values for QS, penumbra and umbra are somewhat lower at 0.26--0.5. The inferred gas densities of 0.75--4.3$\times 10^{-7}$\,g\,cm$^{-3}$ are in the range of the HSRA QS value of 2.89$\times 10^{-7}$\,g\,cm$^{-3}$ used in most of the previous calculations, while for the inner FPs the total pressure and the inferred gas density have similar values as the HSRA at $\log \tau = -0.2$ (Table \ref{table:HSRA}). The magnetic pressure contributes more than 50\,\% in the penumbra and completely dominates the total pressure in the umbra. A similar total pressure just from the gas pressure would only be found for $z<-70$\,km in the HSRA model (Table \ref{table:HSRA}).

\begin{figure*}
\resizebox{17cm}{!}{\includegraphics{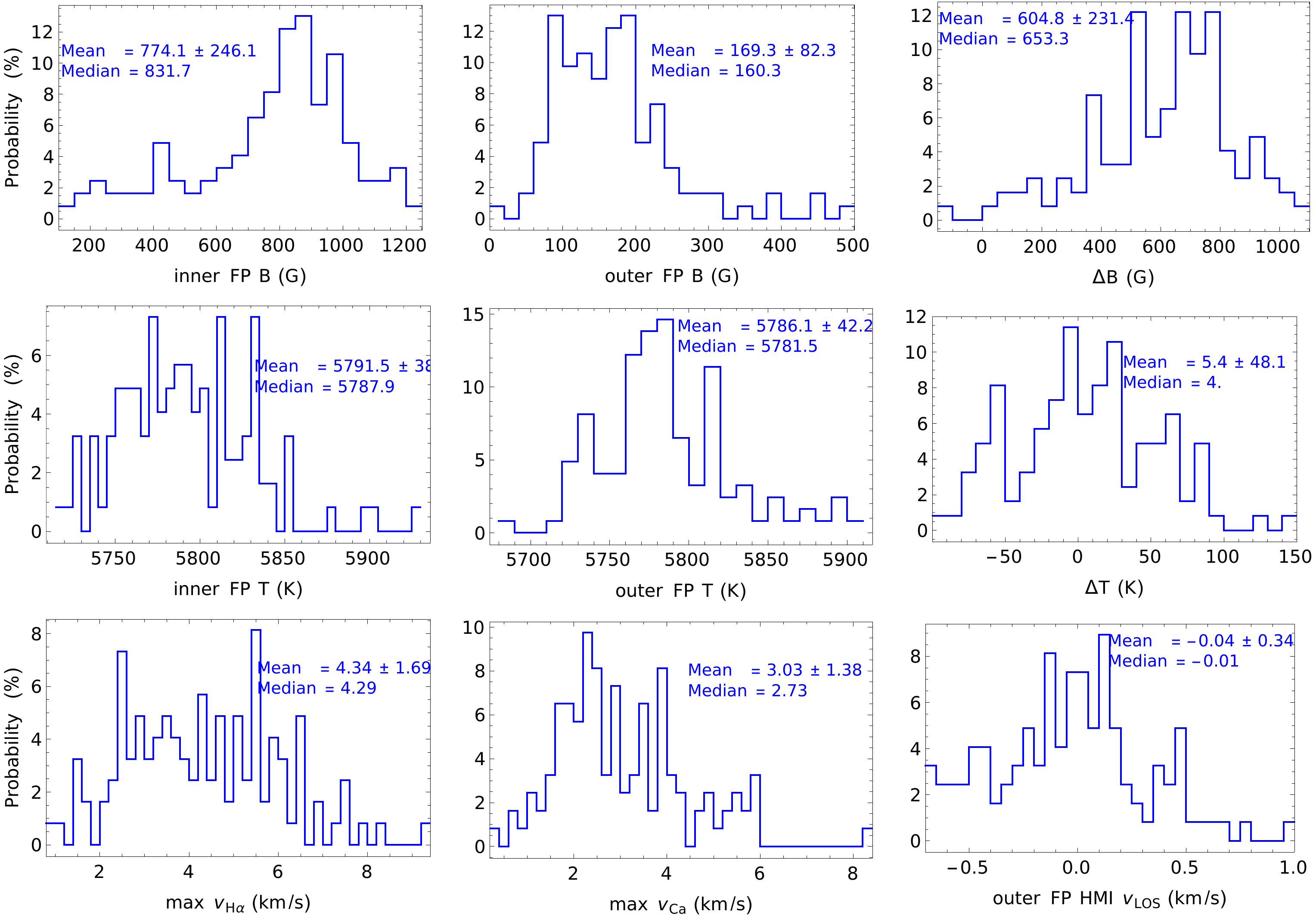}}
\caption{Histograms of physical properties of the manually selected IEF channels. Top two rows: histograms of $B$ and $T$ at the inner (left column) and outer (middle column) FPs and their differences $\Delta B$ and $\Delta T$ (right column). Bottom row, left to right: histograms of maximal chromospheric LOS velocities in H$\alpha$ and \ion{Ca}{ii} IR and photospheric HMI velocity at the outer FPs.}
\label{f:bhist2}
\end{figure*}

\subsection{Manually selected IEF channels}
The manually selected IEF channels are a small subset of the large sample, with seed points for the MFLs chosen on the appearance of IEF channels in the intensity and velocity maps. All parameters related to the magnetic topology (length, height, locations of FPs) were very similar to the average values of the large sample. We only overplotted the histograms for the locations of the inner and outer FPs in Figure \ref{f:distance}. It reveals that the inner FPs of closed MFLs that explicitly correspond to IEF channels are slightly closer to the penumbra than those of the large sample, with about half of them being inside the sunspot. The histograms of field strength and temperature at the inner and outer FPs in the top two rows of Figure \ref{f:bhist2} show that for those MFLs the temperature difference is close to zero ($\Delta T \sim 20$\,K) and negligible compared to the field strength difference ($\Delta B \sim +600$\,G). The chromospheric maximal LOS flow velocities of 3--4\,km\,s$^{-1}$ are slightly higher than the averages for the large sample, while the photospheric HMI velocity at the outer FPs again shows no clear indication of blue shifts or upflows.

\begin{figure}
\resizebox{8cm}{!}{\includegraphics{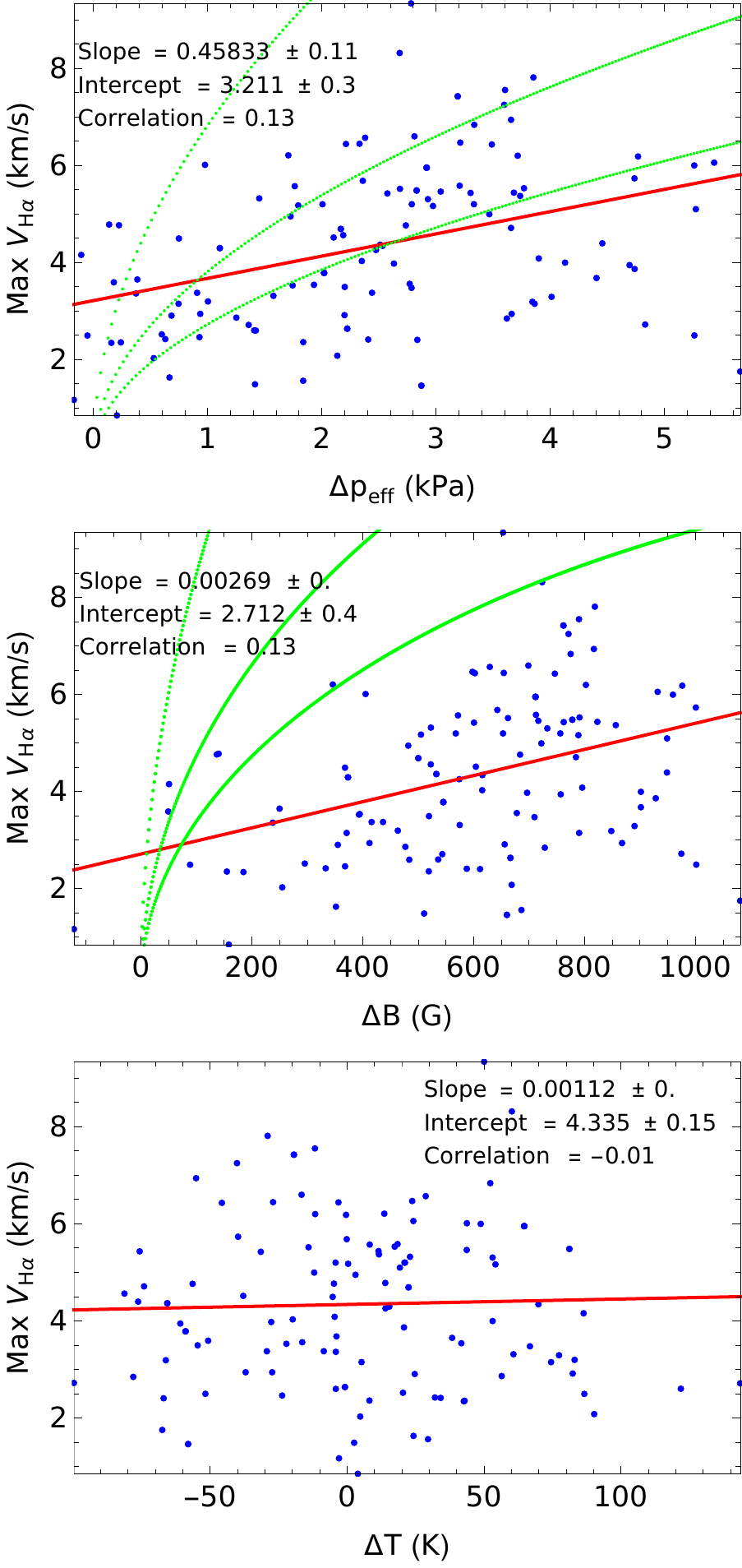}}
\caption{Scatter plots of (top to bottom) $\Delta p$, $\Delta B$ and $\Delta T$ against the maximum H$\alpha$ LOS velocity of manually selected IEF channels. The green lines show the predicted velocities for HSRA gas densities at $\log \tau = -0.3, -1$ and $-2$.}\label{f:scatter2}
\end{figure}

Figure \ref{f:scatter2} shows the observed and predicted velocities as a function of $\Delta p$, $\Delta B$ and $\Delta T$ for the manually selected IEF channels. In that case, the velocities predicted from the effective pressure match the range of observed LOS flow speeds, while they exceed them when only considering the magnetic pressure difference. The temperature difference is primarily positive, which thus slightly reduces the effective pressure difference. The corresponding curve for predicted velocities from $\Delta T$ was below the plot range for $\Delta T > -100$\,K.
\begin{figure}
\resizebox{8cm}{!}{\includegraphics{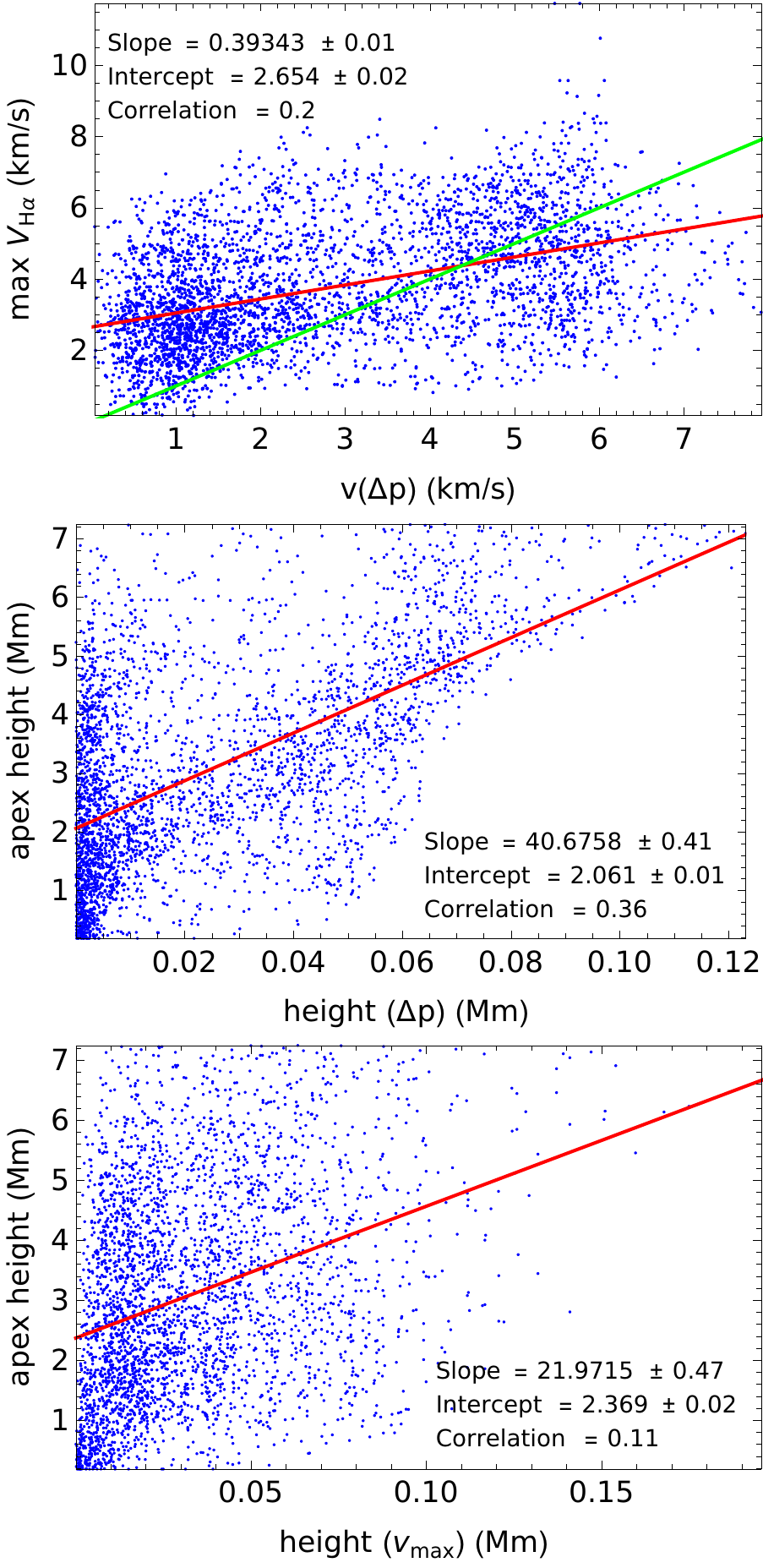}}
\caption{Scatter plots of predicted and measured velocities and apex heights. Top to bottom: predicted vs.~observed maximal H$\alpha$ LOS velocity, measured apex height vs.~prediction from $\Delta p_{\rm eff}$, and measured apex height vs.~prediction from the observed maximal H$\alpha$ LOS velocity. The green line in the top panel indicates a one-to-one correlation.} \label{f:predval}
\end{figure}

\begin{figure*}
\begin{minipage}{8.8cm}
\resizebox{8.8cm}{!}{\includegraphics{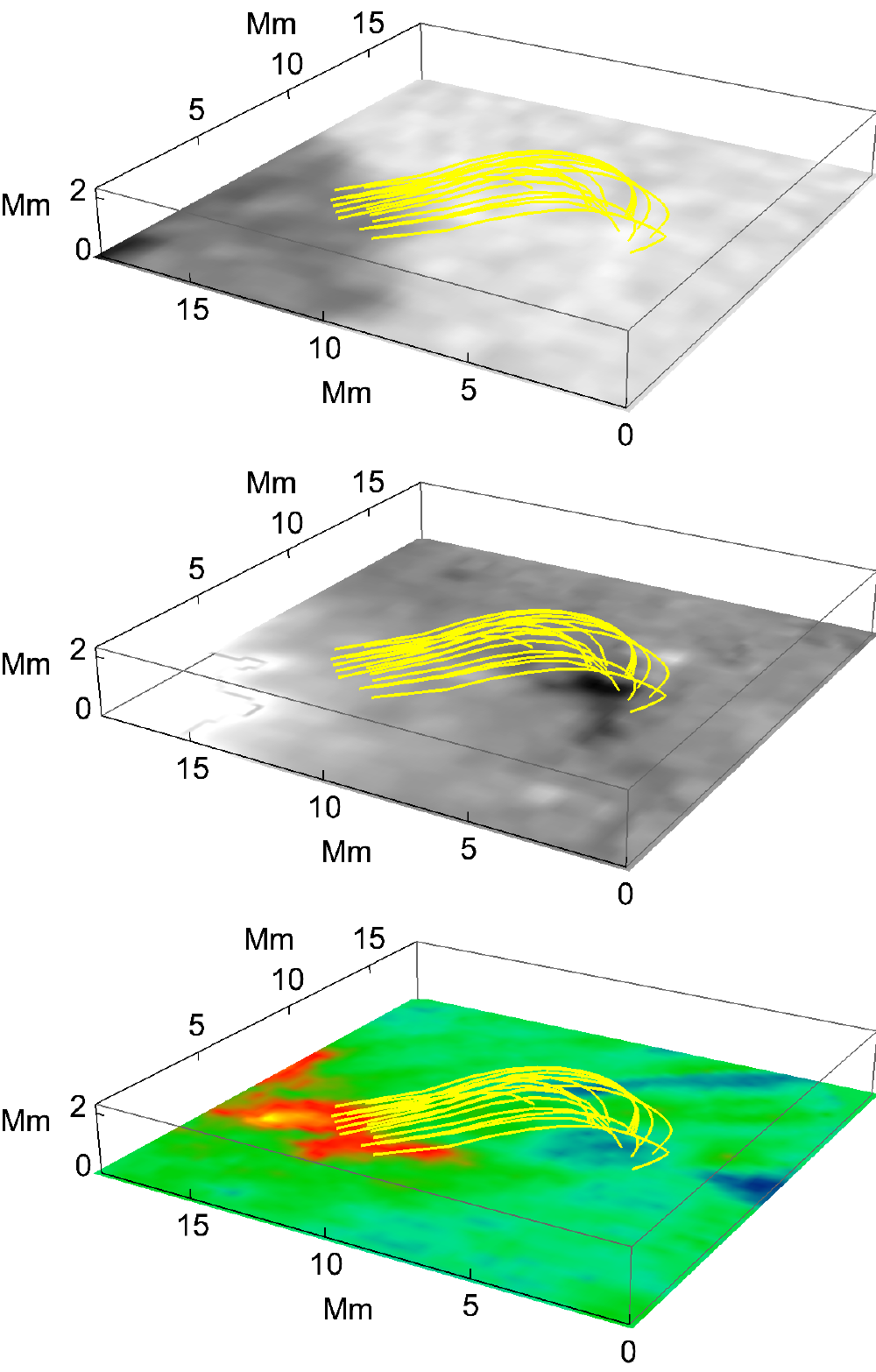}}
\end{minipage}
\begin{minipage}{8.4cm}
\resizebox{8.8cm}{!}{\includegraphics{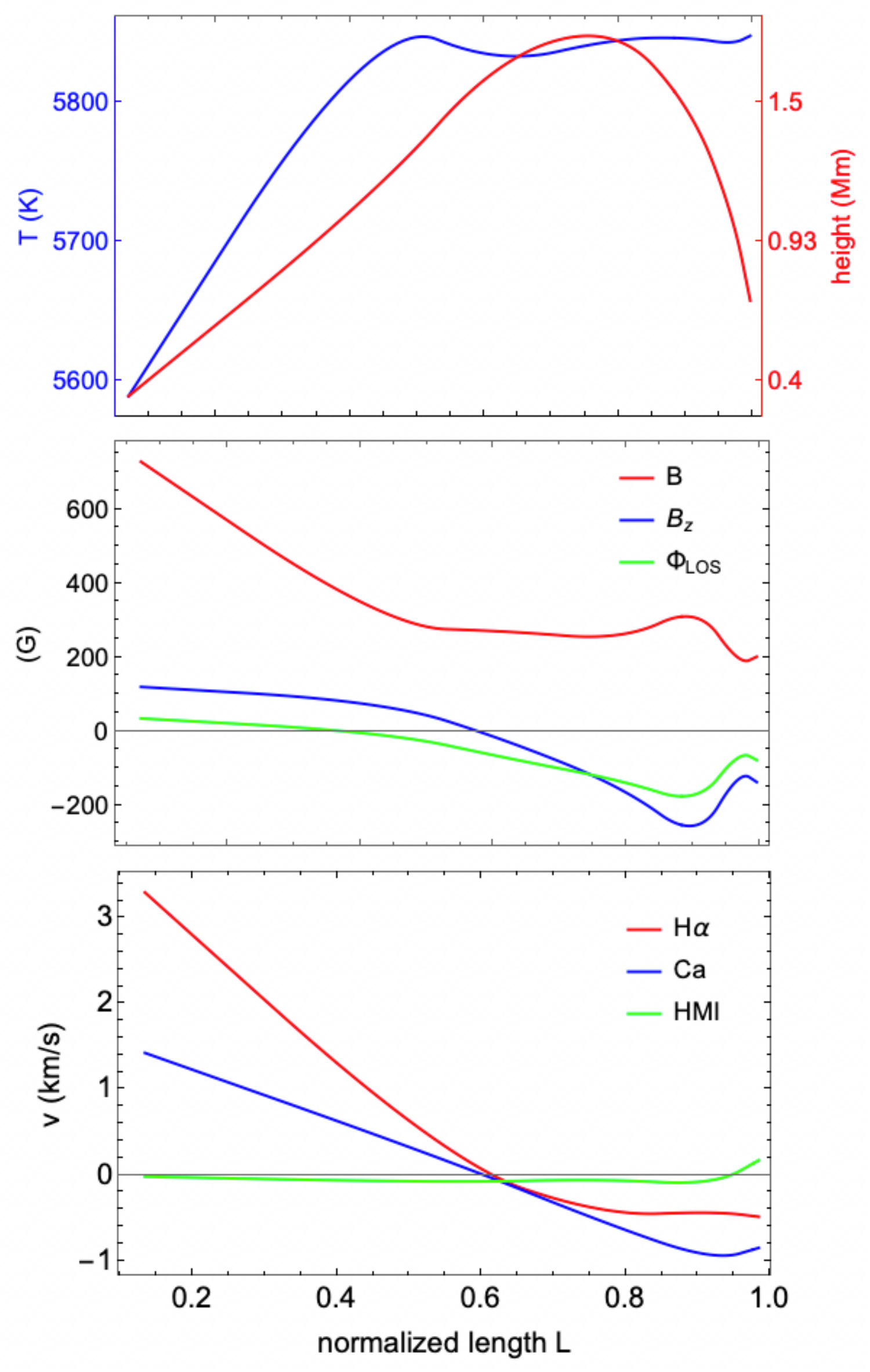}}
\end{minipage}
\caption{The ``perfect" IEF channel south of the sunspot. Left column: 3D view of the MFLs connecting the sunspot to the opposite-polarity patch on top of the continuum intensity $I_c$ (top panel), $B_z$ (middle panel) and the \ion{Ca}{ii} IR LOS velocity (bottom panel). Right column: different quantities averaged over the MFLs in the left column as a function of the relative length $L$. Top panel:  temperature (blue line, y-axis at the left) and height (red line, y-axis at the right). Middle panel: LOS magnetic flux $\Phi_{\rm LOS}$ (green line), $B_z$ (blue line) and field strength $B$ (red line). Bottom panel: LOS velocities of HMI (green line), \ion{Ca}{ii} IR (blue line) and H$\alpha$ (red line).}\label{f:perfectief}
\end{figure*}
\subsection{Predicted flow speeds and apex heights}
Equation (\ref{speed_eq}) allows one to predict the expected flow velocity given the pressure difference, where now a linear relation between $\Delta p$ and $v$ should hold. The top panel of Figure \ref{f:predval} tests the prediction for the maximal H$\alpha$ LOS velocities in the large sample. The correlation coefficient stays comparably low at 0.2, but the data points are to some extent covered by a one-to-one relation. The prediction of the apex height from Equation (\ref{height_eq1}) based on the pressure difference falls short of the apex height measured in the magnetic field extrapolation by a factor of about 50 at a correlation of 0.36 (middle panel of Figure \ref{f:predval}). Apart from being off in magnitude, the trend of the data points follows the prediction, with a complete absence of low measured apex heights for low predicted heights. To some extent, this results from the relation between length and apex height (Figure \ref{f:scatterlength}), where longer MFLs have their inner FPs closer to the umbra which causes a larger $\Delta B$ (Figure \ref{f:neg}), and hence larger $\Delta p$. A similar mismatch by a factor of about 30 in magnitude is seen for the apex height predicted from the observed LOS velocities (bottom panel of Figure \ref{f:predval}).

Assuming a velocity of the order of the chromospheric sound speed of 10\,km\,s$^{-1}$ would only correspond to an apex height of 0.18\,Mm. A pressure difference of 5\,kPa that is covered by the observed values of $\Delta p$ (Figure \ref{f:eff_pressure}) would only be able to lift a gas density of 0.006$\times 10^{-7}$\,g\,cm$^{-3}$ up to the average apex height of 2.96\,Mm. That indicates that the Equations (\ref{height_eq1}) and (\ref{height_eq2}) related to the apex height do not seem to be strictly valid, as flows on the order of 10\,km\,s$^{-1}$ are observed to happen along MFLs of that height.

\subsection{The ``perfect" IEF channel}
To the south-east of the sunspot, one can find several adjacent MFLs (see the square in the rightmost column of Figure \ref{f:selected}) that form an almost ``perfect" IEF channel with all theoretically expected properties : two opposite polarities connected by a low-lying chromospheric loop with blue shifts at the outer and red shifts at the inner end. At the heliocentric angle of the sunspot, this clear velocity signature is rare to find. We will use the average properties of these MFLs to summarize our results, while Table \ref{tab_all} additionally lists average values of different quantities for the large sample and correlation values between them for completeness.

\begin{table*}
\caption{Summary of results for the large sample.}\label{tab_all}
\begin{tabular}{c|c|c|c|c|c|c|c|c}\hline\hline
quantity & apex & 2D $L$ & 3D $L$ & $r_{\rm in}$ & $r_{\rm out}$ & $B_{\rm in}$ & $B_{\rm out}$ & $\Delta B$  \cr\hline
units & \multicolumn{3}{c|}{Mm} & \multicolumn{2}{c|}{$r_0$} & \multicolumn{3}{c}{kG} \cr\hline
$X\pm \sigma_X$ & $3.0\pm1.8$ &$13\pm6$ &$14\pm7$ &$1.2\pm0.2$ &$1.9\pm0.2$ & $0.6\pm0.3$&$0.2\pm0.1$ & $0.4\pm0.3$ \cr\hline\hline
quantity & $T_{\rm in}$ & $T_{\rm out}$ & $\Delta T$ &  LOS $v_{\rm H\alpha}$& Depr $v_{\rm H\alpha}$ &LOS $v_{\rm Ca}$  &Depr $v_{\rm Ca}$  & $v_{\rm HMI}$\cr\hline
 units &\multicolumn{3}{c|}{kK} & \multicolumn{5}{c}{km\,s$^{-1}$}  \cr\hline
$X\pm \sigma_X$  & $5.7\pm0.2$ &$5.8\pm0.03$ & $-0.1\pm 0.2$ &$3.8\pm 1.7$ &$16\pm8$ &$2.5\pm1.2$ & $10\pm5$& $0.1\pm0.3$\cr\hline\hline
quantity & $p_{\rm mag}^{\rm in}$ &  $p_{\rm mag}^{\rm out}$ & $\Delta p_{\rm mag}$ & $p_{\rm gas}^{\rm in}$ &  $p_{\rm gas}^{\rm out}$ & $\Delta p_{\rm gas}$ &  $\Delta p_{\rm eff}$ & $p_{\rm out}^{\rm tot}$  \cr\hline
kPa & $1.6\pm1.7$ & $0.2\pm0.3$ & $1.4\pm1.6$& $10.6\pm0.3$ &$10.8\pm0.1$ & $-0.23\pm0.3$ & $1.67\pm1.9$ & $10.96\pm0.3$ \cr\hline\hline
\multicolumn{9}{c}{Correlation values}\cr\hline\hline
$v_{\rm H\alpha}$-$v_{\rm Ca}$&$L$-apex &$L-\Delta B$ &$\Delta B-\Delta T$ &$\Delta B-v_{\rm H\alpha}$ &$\Delta T-v_{\rm H\alpha}$ & $\Delta p-v_{\rm H\alpha}$ & $T-B^2$& $h(\Delta p)-$ apex\cr\hline
0.66 & 0.90 &0.56 &0.73 & 0.19 &0.17 &0.17 &0.85 & 0.36\cr\hline
\end{tabular}
\end{table*}

The left column of Figure \ref{f:perfectief} shows a magnification of the 3D view of the MFLs on top of the continuum intensity, $B_z$ and the LOS velocity of \ion{Ca}{ii} IR, while the right column shows the corresponding magnetic and thermodynamic quantities along their length averaged over about 50 MFLs. The MFLs connect an opposite-polarity patch of magnetic elements in the sunspot moat at about 15\,Mm distance from the outer penumbral boundary with the latter (left top two panels of Figure \ref{f:perfectief}). The inner FPs of the MFLs are slightly inside the penumbra. The MFLs are close to vertical in the magnetic elements at the outer FPs and more inclined in the sunspot penumbra, forming an arched loop with 2\,Mm apex height (top right panel of Figure \ref{f:perfectief}). The field strength of about 700\,G at the inner FPs is higher than the 300\,G at the outer FPs, which still stand out in $B$, $B_z$ and $\Phi_{\rm LOS}$ relative to their immediate surroundings (middle right panel of Figure \ref{f:perfectief}). The MFLs show in that case clear upflows of 0.5--1\,km\,s$^{-1}$ in H$\alpha$ and \ion{Ca}{ii} IR at the outer and downflows of 1--3\,km\,s$^{-1}$ at the inner FPs (bottom row of Figure \ref{f:perfectief}). A slight blue shift is seen in the photospheric HMI LOS velocities.

For this ``perfect" IEF channel, one thus finds a picture of mass moving upwards at the outer end of an arched magnetic loop that then streams along it into the sunspot. Both the difference in field strength $B$ and temperature $T$ would drive a flow in the same direction. For assumed total pressure balance between the inner and outer FPs, the magnetic, thermal and hydrodynamic topology of these MFLs would all comply with a siphon flow scenario with a mass flow driven primarily by the field strength difference and the inherent gas pressure difference it causes.

\section{Discussion}\label{secdisc}
\subsection{Limitations of the current study}\label{sec_limit}

The current study suffers from a few limitations that to the largest extent are related to or intrinsic to the observational data.

 It could benefit from selecting a variety of sunspots to examine MFL connectivity all around sunspots. For the sunspot used, the IEF channels are best seen on the disk center side of the sunspot, where the LOS and the IEF channels are parallel \citep[Figure \ref{f:ha-dep-vel} and][]{beck+choudhary2019}. The magnetic field extrapolation could not provide closed MFLs in that direction due to the lack of opposite-polarity fields to the west.
 This configuration is similar to that encountered by \citet[][their Figure 12]{kawabata+etal2020}.
 Selecting a leading (trailing) sunspot with trailing (leading) plage to the west (east) of the solar central meridian would improve the situation by likely having more closed MFLs all around the sunspot with our extrapolation technique. A second improvement would be to select a sunspot where the location of the zero line in LOS velocities does not coincide with the inner FPs of the IEF channels on the limb side, which can be achieved by selecting sunspots at small heliocentric angles.

The derivation of the LOS velocities themselves is only possible within limits. As shown in \citet{choudhary+beck2018}, the IEF is only a satellite component in the spectra close to the inner FPs, which leads to an underestimation of the flow velocity by a factor of up to two. The de-projection to the true flow angle and speed was done based on the magnetic field vector at a single height in the NFFF extrapolation, whose accuracy cannot be confirmed with the current data. Additionally, the exact formation heights of the velocities of the chromospheric spectral lines of \ion{Ca}{ii} IR at 854\,nm and especially H$\alpha$ in the canopy of a sunspot off disk center are not well known.

Apart from using the magnetic vector field information as the input for the NFFF extrapolation, the field strength and temperature from the HMI data were directly used. They are to some extent not fully consistent with each other, as their formation heights differ by a few hundred kms \citep{norton+etal2006}. A second limitation is the modulus of the field strength. The HMI magnetograph has only a limited spectral sampling that cannot spectrally resolve the Zeeman splitting for strong magnetic fields. Its spatial sampling of 0\farcs5 is prone to lead to unresolved structures within a single pixel in the QS in addition. The 1-component Milne-Eddington inversion used for the derivation of the vector field from the HMI observations cannot take that into account. The field strength values are thus likely to underestimate the true field strength both in the QS and the sunspot, e.g., the average field strength of 180\,G at the outer FPs (Figure \ref{f:histblarge}) is far from the 1-kG fields expected for magnetic elements in the moat \citep{beck+etal2007,utz+etal2013}.  \citet{beck+choudhary2019} compared the field strength at the inner FPs between HMI and an inversion of the \ion{Fe}{i} lines near 1565\,nm and found the HMI values to be about 500\,G lower than the average value of 1.3 kG from the Fe lines. Even for fields of one kG at the outer FPs, a positive field strength difference of the same modulus as found here (0.3--0.4\,kG) would thus be maintained (Table \ref{tab_all}).

The limitations of the NFFF extrapolation are to some extent caused by the HMI input data. At the photospheric boundary, the values of $B$ are likely off to some extent and the field inclination in the penumbra corresponds to a weighted average in the case of unresolved magnetic field components with different inclinations \citep{bellot+etal2004,beck2008}. The magnetic connectivity is to a largest extent deterministic and can only provide closed MFLs between opposite-polarity FPs, but HMI cannot detect diffuse weak magnetic flux below a certain level because of its spatial and spectral resolution. Photospheric high-resolution observations of internetwork quiet Sun magnetism reveal a large amount of magnetic flux that is beyond HMI's detection capabilities \citep{lites+etal2008,beck+rezaei2009,beck+etal2017}.
A connection between the outer penumbra and weak magnetic flux favors, however, a siphon flow scenario even more because of the resulting small field strength at the outer foot points. The 2D chromospheric flow field of the same sunspot was successfully modeled in \citet{beck+etal2020} assuming an axisymmetric pattern with IEF channels all around the sunspot.

No magnetic field lines match the strong and large filament to the east, whose negative polarity FP could be at about the center of the FOV of the extrapolation box (bottom left panel with the AIA 304\,{\AA} image in Figure \ref{f:nfff}). The magnetic field extrapolation supports here a scenario of a mass loading on top of an arcade of field lines that cross the neutral line. Without introducing an inertial term into it, any extrapolation is, however, not able to yield dips in field lines apart from very peculiar configurations in the input magnetograms.

The energy and pressure balance suffers from all the shortcomings listed above. With the partially unknown formation heights for the LOS velocities, temperature and field strength, it is not obvious which density to attribute to the IEF. The IEF has well-defined inner FPs, where it stops at an optical depth of $\log \tau = -3 \equiv z = 400$\,km \citep{choudhary+beck2018}, but it is not clear at which height it sets in at the outer FPs. We plan to investigate the height variation of the IEF using other data sets with more spectral lines that cover a larger range of formation heights \citep[see, e.g.,][]{felipe+etal2010,bethge+etal2012} in the future.

\subsection{Magnetic properties of the IEF}
Thanks to the results of the NFFF extrapolation, we were able to trace individual MFLs related to the IEF.  We used two different samples, where the large sample was defined based on just the magnetic topology without considering the chromospheric velocities, while the second sample was primarily based on the intensity and velocity signature of IEF  channels. Both samples yielded nearly identical average values and histograms for the properties of the MFLs and their foot points (Figures \ref{f:histblarge}, \ref{f:vel_hist}, \ref{f:bhist2} and \ref{f:scatter2}), which indicates that they trace the same structures in the solar atmosphere, the IEF channels.

The connectivity found from the magnetic field extrapolation in the current study aligns with the results or assumptions on IEF channels in previous studies. The average distances of inner and outer FPs of 1.2 and 1.9 sunspot radii (Figure \ref{f:hist}) implies that they connect the outer penumbra with the end of the moat cell \citep{sobotka+roudier2007}. In \citet{beck+etal2020}, the corresponding values for the same sunspot were 0.98 and 2\,$r_0$, derived or set only using the velocity maps, while peak velocities and intensities for the inner FPs were found at about 1\,$r_0$ in \citet{beck+choudhary2020}.

The average length of closed MFLs of 13\,Mm allows one an indirect estimate of the lifetime of IEF channels. Assuming a motion with the chromospheric sound speed of about 10\,km\,s$^{-1}$, it takes about 20\,min to reach the inner FP from the outer end. Typical life times of IEF channels then should be of the same order when an IEF channels is established, which matches the life times of 10--60\,min found in \citet{beck+choudhary2020} from tracing the flow signature of IEF channels with time.

The current average apex height of 3\,Mm matches the one of 2.45\,Mm inferred for this sunspot based only on the LOS velocity maps in \citet[][]{beck+etal2020} quite closely.
\citet{aschwanden+etal2016} found a typical height of up to 4\,Mm for chromospheric structures such as superpenumbral fibrils, with a similar steep decline of the histogram towards larger heights (Figure \ref{f:hist} and their Figure 10). The assumed parabolic loop shape used in \citet{beck+etal2020}, which was previously also used in \citet{maltby1975}, is directly confirmed to first order by the field extrapolation (Figure \ref{f:scatterlength}), which also supports the inclination values for these MFLs being around $\pm 30^\circ$ to the local horizontal \citep{haugen1969,dialetis+etal1985,beck+choudhary2019,beck+etal2020} with a smooth radial variation. The comparably solid ratio between length and height of 4:1 (Figure \ref{f:scatterlength}) might allow one to at least get some estimate of apex heights for closed MFLs in the absence of an extrapolation when the horizontal foot point distance is known.

The average field strengths at the inner and outer FPs of 0.6\,kG and 0.2\,kG in the HMI data are likely underestimating the true values,
 but the average positive field strength difference of +0.4\,kG is, however, solid in its sign. One of the FPs of the closed MFLs related to the IEF is inside or close to the outer boundary of the penumbra, which implies that nearly any connection will end in an outer FP of lower field strength. The properties of the outer FPs in location, field strength and inclination (Figures \ref{f:shape}, \ref{f:footpoints}, \ref{f:bshape} and \ref{f:perfectief}) match to those of magnetic elements in the QS with a higher field strength than their immediate surroundings and nearly vertical magnetic field inclinations. The  difference of the temperature of the inner and outer FPs seems to play a smaller role than $\Delta B$ in most cases, with the caveat that the HMI resolution might prevent to see the true temperatures of the intergranular lanes in which the magnetic elements of the outer foot points are likely to reside.

\subsection{Pressure balance}
The simplified pressure balance of Equation (\ref{eq_inoutp}) together with the conversion to the expected flow speed of Equation (\ref{speed_eq}) correctly predicts the square-root shape of the distribution of the $\Delta p - v$ scatter plots, but the numbers line only partly up with the observed LOS or de-projected velocity values for individual IEF channels (Figure \ref{f:predval}). The main reasons for this will be the observational limitations discussed in Section \ref{sec_limit}. Varying the gas density used in the calculations within reasonable limits ($z =20-200$\,km, $\rho = 0.3-3 \times 10^{-7}$\,g\,cm$^{-3}$) suffices to cover the range of observed LOS velocities of $2-10$\,km\,s$^{-1}$.

For the umbra and the QS with primarily vertical magnetic fields and lower higher-order magnetic contributions such as curvature forces to the pressure equation \citep{steiner+etal1986,borrero+etal2019}, the relation between $T$ and $B^2$ (Figure \ref{f:tvsb}) might allow one to derive not only average density values, but also spatially resolved density maps when a suited boundary value for the total pressure is prescribed and the Wilson depression is properly accounted for \citep{puschmann+etal2010,loeptien+etal2020}. A combination of the magnetic field extrapolation results with a thermal inversion of the \ion{Ca}{ii} IR spectra, e.g., through the Calcium Inversion based on a Spectral Archive code \citep[CAISAR;][]{beck+etal2013,beck+etal2015}, would possibly provide density stratification in a 3D volume.

\subsection{The driver of the IEF}
The IEF channels are well aligned with the magnetic field vector near the inner FPs \citep{beck+choudhary2019}. Their visibility in especially the \ion{Ca}{ii} IR data (Figures \ref{f:full}, \ref{f:selected} and \ref{f:perfectief}) complies with the shape in the magnetic field extrapolation (Figure \ref{f:shape}) that predicts that the IEF channels leave the \ion{Ca}{ii} IR formation height of 1--2\,Mm around the apex. The  primarily radial orientation of the MFLs matches that of the intensity fibrils in H$\alpha$  or \ion{Ca}{ii} IR in Figure \ref{f:full}  at most places \citep[see also][]{delacruz+socasnavarro2011,schad+etal2013,leenaarts+etal2015,kawabata+etal2020} apart from the dark filaments that supposedly do not correspond to IEF channels. There is thus no indication that the IEF would not follow the closed MFLs of the NFFF extrapolation wherever the flow cannot be continuously traced in the velocity maps, or that no IEF channels with similarly shaped closed field lines would exist where the magnetic field extrapolation is unable to yield them. With the positive field strength difference and the anti-correlated temperature difference towards the inner FPs, a siphon flow is therefore the most likely explanation for the IEF pattern around sunspots.

The diagnosis of the driver of the IEF in this study was only possible through the combination of high-resolution chromospheric spectra and velocities -- or high-resolution observations in general -- with a magnetic field extrapolation derived from full-disk data of lower spatial resolution. It highlights the potential of magnetic field extrapolations for the correct interpretation of the physics visible in high-resolution observations at much smaller spatial scales. This approach has a broad range of potential applications that have only partially been explored yet \citep[e.g.,][]{aschwanden+etal2016,grant+etal2018,yadav+etal2019,kawabata+etal2020,louis+2021} and could be developed into a standard diagnostic tool for high-resolution observations in the future. The availability of a magnetic field extrapolation provides also additional diagnostic potential through, for instance, the de-projection of LOS velocities to the field direction to obtain true flow speeds. In the other direction, high-resolution observations could help in detailed modeling of, e.g., filaments by determining a corresponding mass density from H$\alpha$ spectra to add as a constraint to the extrapolation or the extrapolation results. Combinations of these two different approaches have a clear potential to improve future research in solar physics, where the list of techniques applied in the current study is far from being exhaustive.
\section{Conclusions}\label{secconcl}
We find that the inverse Evershed flow of the leading sunspot in AR NOAA 12418 happens along elongated arched loops of about 13\,Mm length with an apex height of about 3\,Mm. The corresponding closed magnetic field lines connect the outer penumbra with magnetic elements in or at the end of the moat cell. The positive difference in magnetic field strength of on average +400\,G and the negative one in temperature of -100\,K both support a siphon flow from the outer foot points towards the sunspot as the driver of the inverse Evershed flow.

\begin{acknowledgements}
The Dunn Solar Telescope at Sacramento Peak/NM was operated by the National Solar Observatory (NSO). NSO is operated by the Association of Universities for Research in Astronomy (AURA), Inc. under cooperative agreement with the National Science Foundation (NSF). HMI data are courtesy of NASA/SDO and the HMI science team. IBIS has been designed and constructed by the INAF/Osservatorio Astrofisico di Arcetri with contributions from the Universit{\`a} di Firenze, the Universit{\`a}di Roma Tor Vergata, and upgraded with further contributions from NSO and Queens University Belfast. This work was supported through NSF grants AGS-1413686 and AGS-2050340. Q.H. and A.P. acknowledge partial support of NASA grants 80NSSC17K0016, LWS 80NSSC21K0003 and NSF awards AGS-1650854 and AGS-1954503.This research was also supported by the Research Council of Norway through its Centres of Excellence scheme, project number 262622, as well as through the Synergy Grant number 810218 459 (ERC-2018-SyG) of the European Research Council.
\end{acknowledgements}

   \bibliographystyle{aa} 
   \bibliography{ief} 
%
\begin{appendix} 
\section{Derivation and calibration of line-of-sight velocities}\label{appendix_a}
\subsection{Photospheric velocities from HMI}\label{app_photvelo} The sunspot of AR NOAA 12418 was located at 35.54$^\circ$ eastern longitude.
The average LOS velocity caused by solar rotation was $-1$\,km\,s$^{-1}$ with an additional variation of 0.4\,km\,s$^{-1}$ across the box used for the magnetic extrapolation (middle column of Figure \ref{fig_overview}), which required to use a spatially varying correction as well.
We determined the contribution of the solar rotation by averaging the de-projected (see Appendix \ref{deproject_fulldisk} below) HMI velocity maps in the extrapolation box in the $x$ and $y$ directions and fitted straight lines to the average curves. The fitted straight lines were then subtracted from the data. After subtraction of these straight lines, the average HMI velocity in QS regions was about zero. \citet{loehner+etal2019} measured a convective blue shift of about $-0.2$\,km\,s$^{-1}$ for the HMI \ion{Fe}{i} line at 617.3\,nm at $\mu=0.7$. We did not apply an additional correction for that, but note that HMI velocities of about zero should correspond to small blue shifts of -0.2\,km\,s$^{-1}$.

\subsection{Bisector analysis of chromospheric spectra}\label{app_chromvelo} 
\begin{figure}
\resizebox{8.8cm}{!}{\includegraphics{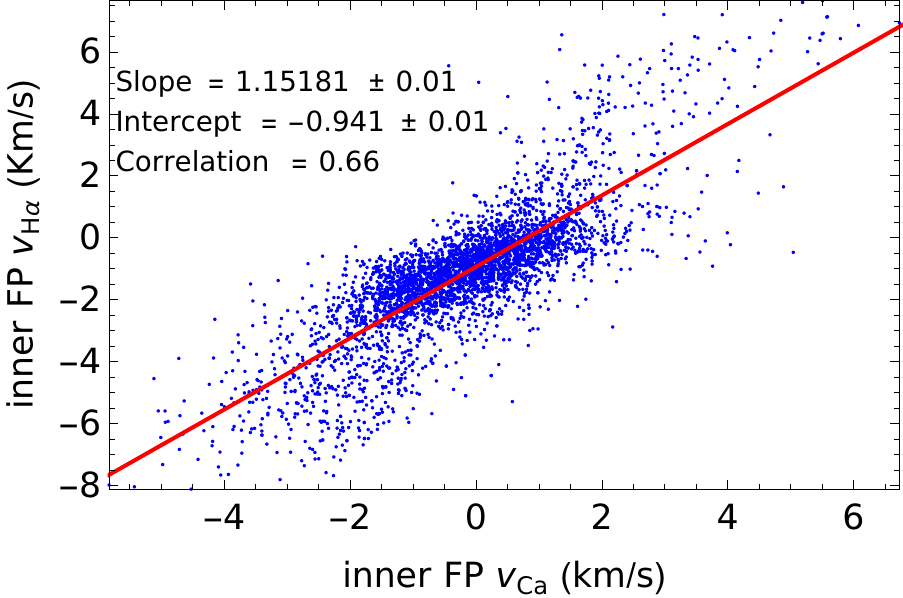}}
\caption{Scatter plot of the LOS velocities of H$\alpha$ and \ion{Ca}{ii} IR at the inner FPs of closed MFLs near the outer penumbral boundary. The red line shows a linear regression to the data points. Its slope and intercept, and the linear correlation coefficient are given inside the panel. Only every 5th data point is plotted.}
\label{f:ca-ha-vel}
\end{figure}

We used the same bisector analysis as in \citet{beck+choudhary2020} and selected the bisector velocity at 93\,\% line depth as LOS velocity of \ion{Ca}{ii} IR and H$\alpha$. The average velocity across the whole IBIS FOV was set to zero for each of the six spectral scans, which puts the average umbral velocity at rest as well \citep{beck+etal2020,henriques+etal2020}. The velocities of H$\alpha$ and \ion{Ca}{ii} IR in and near sunspots are similar \citep[e.g.,][]{beck+choudhary2020,beck+etal2020}, with a correlation coefficient above 0.5 and usually higher velocity values in H$\alpha$ (Figure \ref{f:ca-ha-vel}). We will primarily show results for H$\alpha$ in the following but note that the velocities of the two lines can be used synonymously. The line-core intensity was defined as the lowest value inside the spectral line in each profile.

\section{Conversion of HMI intensity to temperature}\label{appendix_b}
We used the Planck function to convert the de-projected continuum intensity maps of HMI to temperature \citep[see, e.g.,][]{beck+etal2012,valio+etal2020}. \citet{norton+etal2006} give the formation height of the continuum (line core) of the \ion{Fe}{i} line at 617\,nm as about 20\,km (300\,km), so the assumption of an ideal gas in local thermodynamic equilibrium (LTE) with black body radiation is justified. At the heliocentric angle of the observations, the continuum formation height should shift slightly upwards to about 40\,km 
with the optical depth changing from $\tau = 0.63$ to $\tau = 0.63\,\sin\,\theta = 0.46$ corresponding to $\log \tau = -0.2$ and $\log \tau = -0.33$, respectively.

We calculated the emergent radiation at 617\,nm from the Planck function for a temperature range from  3000 to 7000\,K  and normalized the values by the intensity for 5800\,K to obtain a conversion curve from relative intensities to absolute temperatures. The center-to-limb variation (CLV) in the observed intensity maps was removed in a similar way as for the HMI velocities using averages in $x$ and $y$, but now dividing with the fitted straight lines. After the correction for the CLV, the relative HMI intensities were converted to their corresponding temperatures. With this choice of normalization, the average QS temperature is forced to be 5800\,K.

\begin{figure}
\resizebox{8.8cm}{!}{\includegraphics{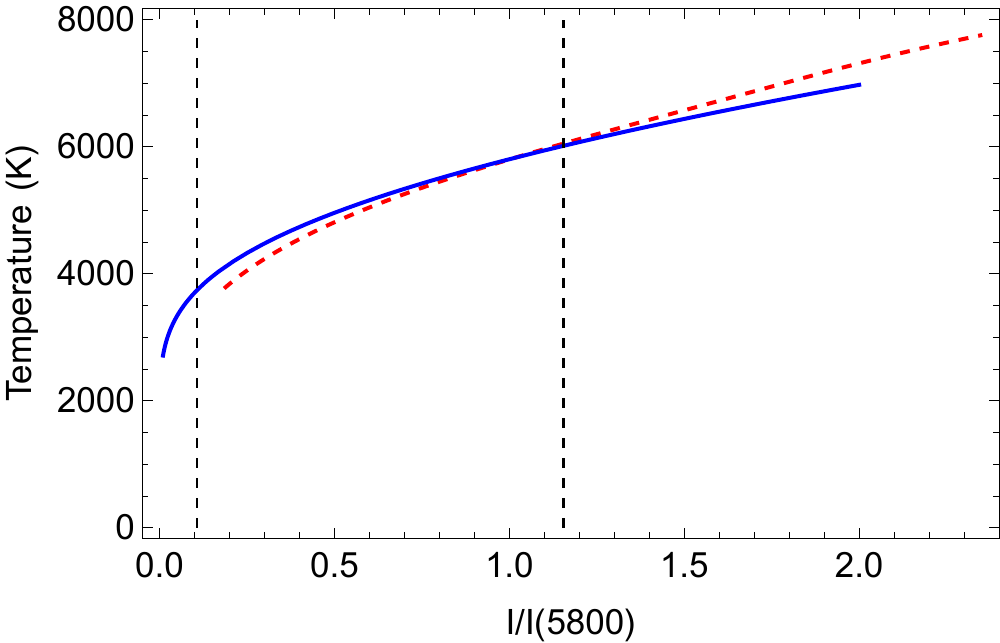}}
\caption{Conversion curves between HMI continuum intensity and temperature as derived from the Planck function (blue line) and using the SIR code (red dashed line). The black dashed vertical lines at $I/I(5800)$ = 0.11 and 1.15 represent the minimum and maximum intensity values in our HMI data.}
\label{f:planck-sir}
\end{figure}

 To verify the conversion curve, we generated the same using the Stokes Inversion based on Response functions code \citep[SIR;][]{ruizcobo+deltoroiniesta1992} that employs LTE. We synthesized a wavelength window around the \ion{Fe}{i} line at 617.3\,nm that covered continuum wavelengths while modifying the HSRA model with global offsets of -2000 to +2000\,K at all optical depths. We selected the continuum intensity that corresponded to having a temperature of 5765\,K at $\log \tau = -0.4$ in the original unperturbed HSRA model for the normalization in that case. Figure \ref{f:planck-sir} demonstrates that in the relevant intensity range of the HMI data from 0.11 to 1.15 the differences between the two approaches are minor, so the simple conversion using the Planck function can well be used, especially since our main focus is on spatial locations with a relative intensity of about unity.

\section{Non-force-free magnetic field extrapolation}\label{appendix_c}
We used the NFFF extrapolation code developed by \citet{hu10} to obtain the magnetic field connectivity around the sunspot. The algorithm for this code is based on the principle of minimum dissipation rate \citep[MDR;][]{montgomery88,dasgupta98,bhattacharyya04}, which originates from a variation approach that allows one to obtain dissipative relaxed states in a two-fluid plasma with an external helicity driving \citep{bhattacharyya07}. This thus makes it suitable for an open system with a flow like the solar photosphere.

The MDR approach leads to a set of two decoupled inhomogeneous double-curl Beltrami equations \citep{mahajan98} for the magnetic field $\mathbf{B}$ and the fluid vorticity $\boldsymbol{\omega}$
given by \citep{bhattacharyya04,bhattacharyya07}

\begin{align}
\nabla\times\nabla\times\mathbf{B}+a_1\nabla\times\mathbf{B}+b_1\mathbf{B}&=\nabla \psi\label{e:bnff}\\
\nabla\times\nabla\times\boldsymbol{\omega}+a_2\nabla\times\boldsymbol{\omega}+b_2\boldsymbol{\omega}&=\nabla \chi\,,
\end{align}

where $a_1, a_2, b_1$, and $b_2$ are constants that depend on the parameters of the system, and $\psi$ and $\chi$ are arbitrary scalar functions that satisfy the Laplace's equation.

For the rest of the description, we focus only on the magnetic field, which is more relevant to the current study. The ambiguity arising from the arbitrary potential can be eliminated by taking the curl of Equation \eqref{e:bnff}, which results in \citep{hu08b}

\begin{equation}
\nabla\times\nabla\times\nabla\times\mathbf{B}+a_1\nabla\times\nabla\times\mathbf{B}+b_1\nabla\times\mathbf{B}=0\,.\label{e:bnff3}
\end{equation}

An exact solution of Equation \eqref{e:bnff3} can be obtained using the linear superposition of three linear force-free fields (LFFFs), arising from the orthogonality of Chandrasekhar-Kendall (CK) eigenfunctions \citep{chandrasekhar57}. Thus, $\mathbf{B}$ is expressed as \citep{hu08b}:

\begin{equation}
\mathbf{B} = \mathbf{B_1}+\mathbf{B_2}+\mathbf{B_3}; \quad \nabla\times\mathbf{B_i}=\alpha_i\mathbf{B_i}\,,
\label{e:b123}
\end{equation}

where $\alpha_i$ are distinct constant parameters with $i=1,2,3$. The equation further requires that one of the $\alpha_i$ is zero. Here we arbitrarily choose $\alpha_2=0$, which makes $\mathbf{B_2}$ a potential field. This then implies that $a_1=-(\alpha_1+\alpha_3)$ and $b_1=\alpha_1 \alpha_3$.
Now, we combine Equations  \eqref{e:bnff3} and \eqref{e:b123} to get

\begin{align}
\begin{pmatrix}
\mathbf{B_1}\\
\mathbf{B_2}\\
\mathbf{B_3}\\
\end{pmatrix}
=\mathcal{V}^{-1}
\begin{pmatrix}
\mathbf{B}\\
\nabla\times\mathbf{B}\\
\nabla\times\nabla\times\mathbf{B}\\
\end{pmatrix}\,.
\label{e:vmat}
\end{align}

Here $\mathcal{V}$ is called the Vandermonde matrix whose elements are of the form $\alpha^{i-1}j$ for $i, j = 1, 2, 3$ \citep{hu08a}.
Writing the above equation for the $z$ component yields the boundary condition for each LFFF. Assuming a value for the $\alpha_i$ parameter, we can then use a standard fast Fourier transform based LFFF solver \citep{alissandrakis81} to obtain the extrapolated field in the full volume.

To obtain an optimal pair of ($\alpha_1, \alpha_3$), we minimize the difference between the observed ($\mathbf{B}_t$) and computed ($\mathbf{b}_t$) transverse field by defining the following metric \citep{hu08a,hu08b}:

\begin{equation}
E_n =\sum_{i=1}^M |\mathbf{B}_{t,i}-\mathbf{b}_{t,i}|/\sum_{i=1}^M |\mathbf{B}_{t,i}|\,.
\end{equation}

Here $M=N^2$, is the total number of grids points on the bottom boundary.

One term on the right-hand side of Equation \eqref{e:vmat} involves the evaluation of a second derivative, $(\nabla\times\nabla\times\mathbf{B})_z=-\nabla^2 B_z$, at $z=0$. This means that we need to provide the boundary conditions at two or more layers. Since vector magnetograms are available only at the photospheric boundary, \citet{hu10} devised an iterative scheme which successively corrects the potential subfield $\mathbf{B_2}$ starting from an initial guess. By setting $\mathbf{B_2}=0$, Equation \eqref{e:vmat} is first reduced to a 2nd-order matrix equation. This allows us to unambiguously determine the boundary conditions for subfields $\mathbf{B_1}$ and $\mathbf{B_3}$. If the value of the metric $E_n$ is above a certain threshold, then a corrector potential field, which is derived from the difference in the observed and the computed transverse field, is added to $\mathbf{B_2}$ to improve the agreement. For the extrapolations in the present study, the final value of the $E_n$ was around 0.32, which is similar to those obtained in previous studies \citep{liu+2020ApJ}.

\section{De-projection and alignment of high-resolution and full-disk data}\label{appendix_d}

\subsection{Spatial De-projection of High-Resolution Data}
\begin{figure}
\resizebox{8.8cm}{!}{\includegraphics{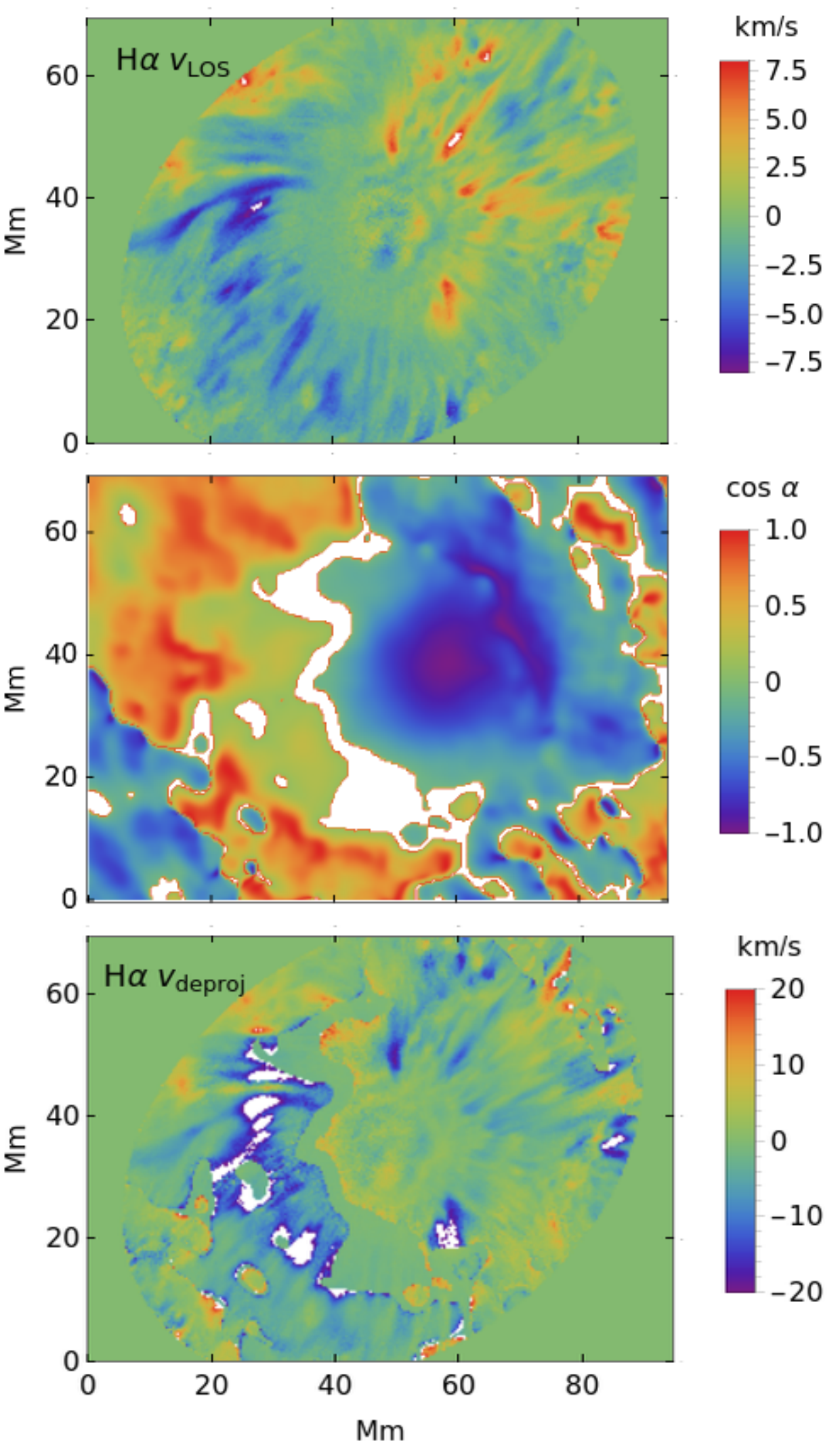}}
\caption{De-projection of LOS velocities onto the magnetic field vector. Top to bottom: observed H$\alpha$ LOS velocity, $\cos \alpha (x,y)$, and de-projected H$\alpha$ velocities. No de-projection was applied for all places with $\cos \alpha <0.1$ and $|v| < 0.2\,$km\,s$^{-1}$ (white areas in the middle panel).}
\label{f:ha-dep-vel}
\end{figure}
Because of the off-center position of the AR, the high-resolution data from the DST required a de-projection to compensate their geometrical foreshortening. We first determined the center of the sunspot and the orientation of the line that joins the centers of the sunspot and the sun, which made an angle of 35.54$^\circ$ to solar east-west. We rotated all 2D images by the angle above to have the symmetry line along a row of the data. We then stretched the images by a factor of $1/\cos \theta = 1.37$ with $\theta = 43^{\circ}$ being the heliocentric angle of the sunspot in the y-axis and rotated the images back by 35.54$^\circ$ to their original north-south orientation. The resulting images now had 1365$\times$1000 pixels instead of the initial 1000$\times$1000 pixels in the uncorrected DST images. The 1-hr duration of the observation was too short to cause a significant change in the position of the sunspot on the disk.  Hence, the same values for the angle and stretching were used for the all the IBIS images recorded during the observation.

\subsection{De-projection of LOS velocities}\label{app_deprovelo}

The observed velocities are the projection of the true velocity vector onto the LOS. The true flow speed can only be determined if the flow angle is known, which cannot be directly derived from the observed spectra, only through additional assumptions such as axisymmetry \citep{schlichenmaier+schmidt2000} or calculations such as a thermal inversion \citep{beck+choudhary2019}. A third possibility is to assume field-aligned flows \citep[e.g.,][]{bellot+balthasar+etal2003,beck2008}, where ionized and magnetized plasma can only move along MFLs but not perpendicular to them. In the vicinity of a sunspot at chromospheric layers this assumption is largely valid. We thus used the magnetic field extrapolation to retrieve the magnetic field vector ${\bf B}(x,y,z)$ at a height $z=1$\,Mm across the FOV. The height corresponds roughly to the formation height of the \ion{Ca}{ii} IR line core in which the IEF channels can be seen both in intensity and velocity maps. The scalar product of the LOS vector $\vec{\bf LOS} = [\cos \beta \, \sin \theta ,\sin \beta \sin \theta,-\cos \theta]$ with $\beta = 215.54^\circ$ and $\theta = 43^\circ$ with the magnetic field vector gives

\begin{equation}
\vec{\bf LOS} \cdot {\bf B}(x,y) = |B|\,\cos \alpha (x,y)\,,
\end{equation}

while the de-projected true flow speed is given by $v_{\rm de-proj}(x,y) = v_{LOS}(x,y)/ \cos \alpha(x,y)$.

Figure \ref{f:ha-dep-vel} shows an example of the H$\alpha$ LOS velocities prior and after the de-projection for one velocity map together with the corresponding map of $\cos \alpha$. On the center side, the LOS is roughly aligned with the magnetic field, so there is little change to the flow speed apart from the sign. On the limb side, there is some extended area where the LOS was perpendicular to the magnetic field and the values of both the observed LOS velocities and of $\cos \alpha$ are very small. We decided not to apply the de-projection to all pixels $(x,y)$ with $|\cos \alpha| < 0.1$ or an unsigned LOS velocity $|v| < 0.2$\,km\,s$^{-1}$ to avoid introducing spurious high velocities, but set $\cos \alpha$ to 1 and the LOS velocities to zero at those places. With the lower threshold of 0.1, the de-projection thus increased the LOS flow speeds by a factor of 1--10 depending on the location in the FOV.
\subsection{De-projection and mapping of full-disk data}\label{deproject_fulldisk}
For the NFFF extrapolations, we used the ``hmi.sharp\_cea\_720s" data series from HMI, which provides the three components of the magnetic field remapped onto a heliographic cylindrical equal-area (CEA) coordinate system centered on an active region cutout \citep{bobra+2014soph}.
The original FOV downloaded from JSOC\footnote{http://jsoc.stanford.edu/ajax/lookdata.html} consisted of 1089$\times$541 pixels centered at 192.51\degree~and -15.58\degree~Carrington longitude and latitude, respectively. This was then cropped to 1024$\times$512 pixels by shifting the origin to (60,10) pixels to improve the numerical accuracy. The full 3D domain for the extrapolation then consists of $1024\times 512 \times 512$ ~pixels in the $x$, $y$ and $z$ directions, respectively. With the 0\farcs5 pixel$^{-1}$ sampling of HMI, the horizontal extent of the box in $x$ is $\sim$371 Mm. All other SDO/AIA filtergrams were also CEA projected and remapped to the same spatial sampling as the magnetic field data with the same FOV.
 \subsection{Alignment of high-resolution and full-disk data}
The de-projected IBIS data (1365$\times$1000 pixels) were degraded to the HMI sampling of 0\farcs5 pixel$^{-1}$ (262$\times$192 pixels). The resampled IBIS data were then placed into an empty array of 1024$\times $512 pixels that matched the size of the FOV used in the extrapolation. The appropriate location for the IBIS images was determined from a visual comparison of the outer penumbral contour line in the HMI and IBIS continuum intensity images. The final images have a common spatial $(x,y)$ coordinate system for all quantities and are aligned to pixel precision at the HMI spatial sampling (rightmost column of Figure \ref{fig_overview}).



\end{appendix}
\end{document}